\documentclass[%
 aip,
 amsmath,amssymb,
 reprint,%
]{revtex4-1}

\usepackage{graphicx}
\usepackage{dcolumn}
\usepackage{bm}

\usepackage[utf8]{inputenc}
\usepackage[T1]{fontenc}
\usepackage{mathptmx}

\begin{document}


\title{Synchronization of clocks and metronomes: \\ A perturbation analysis based on multiple timescales} 




\author{Guillermo H Goldsztein}
\email{ggold@math.gatech.edu}
\affiliation{School of Mathematics, Georgia Institute of Technology, Atlanta, Georgia 30332, USA}
\author{Alice N Nadeau}
\email{a.nadeau@cornell.edu}
\affiliation{Department of Mathematics, Cornell University, Ithaca, NY 14853, USA}
\author{Steven H Strogatz}
\email{strogatz@cornell.edu}
\affiliation{Department of Mathematics, Cornell University, Ithaca, NY 14853, USA}


\date{\today}

\begin{abstract}
In 1665, Huygens observed that two pendulum clocks hanging from the same board became synchronized in antiphase after hundreds of swings. On the other hand, modern experiments with metronomes placed on a movable platform show that they often tend to synchronize in phase, not antiphase. Here we study both in-phase and antiphase synchronization in a model of pendulum clocks and metronomes and analyze their long-term dynamics  with the tools of perturbation theory.  Specifically, we exploit the separation of timescales between the fast oscillations of the individual pendulums and the much slower adjustments of their amplitudes and phases. By scaling the equations appropriately and applying the method of multiple timescales, we derive explicit formulas for the regimes in parameter space where either antiphase or in-phase synchronization are stable, or where both are stable. Although this sort of perturbative analysis is standard in other parts of nonlinear science, it has been applied surprisingly rarely in the context of Huygens's clocks. Unusual features of our approach include its treatment of the escapement mechanism, a small-angle approximation up to cubic order, and both a two- and three-timescale asymptotic analysis.
\end{abstract}


\maketitle 

\begin{quotation}
The ``sympathy of clocks'' that Huygens discovered more than 350 years ago continues to fascinate scientists and lay people alike. 
Although many researchers have shed light on this synchronization phenomenon with a  variety of experimental, analytical, and computational techniques, questions remain about what exactly causes a pair of pendulum clocks to get in sync with one another. 
Adding to the puzzle, a related system -- a pair of metronomes placed on a platform that can move from side to side -- has generated widespread interest. As seen by millions of viewers on YouTube, such metronomes tend to fall into sync spontaneously, but unlike Huygens's clocks, they usually end up with their arms swinging in the \emph{same} direction (in phase), rather than in \emph{opposite} directions (in antiphase). Here, we explore the factors that favor  in-phase or antiphase synchronization, or that allow them both to coexist. We consider a  mathematical model applicable to both pendulum clocks and metronomes and use perturbation theory to predict its long-term behavior. For example, we predict that a pair of identical metronomes on a moving platform will  synchronize in phase if the platform is sufficiently lightly damped. At intermediate damping, both modes of synchronization become possible, depending on initial conditions. And  at sufficiently high damping, only antiphase synchronization is stable. These predictions agree with previous experimental results. More broadly, our perturbative approach is  flexible enough to accommodate many of the variants of Huygens's clocks that have been studied experimentally, while providing a useful tool for their analysis. 
%
\end{quotation}

\section{Introduction}
Synchronization occurs in diverse physical, biological, and chemical systems~\cite{Winfree1980geometry, pikovsky2001synchronization, strogatz2003sync,blekhmansynchronization}. Examples include the synchronous flashing of fireflies, the chorusing of crickets, the rhythmic applause of concert audiences, the coordinated beating of cardiac pacemaker cells, the pathological neural synchrony associated with epileptic seizures, and the coherent voltage oscillations of superconducting Josephson junction arrays.

\begin{figure}
\centering
\includegraphics[width=2.2in]{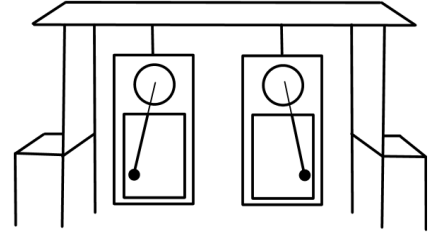}
\caption{\label{fig1}Antiphase synchronization of two pendulum clocks.}
\end{figure}

Historically, the study of synchronization began with Christiaan Huygens's  discovery of an effect he described as ``marvelous''~\cite{huygens1893oeuvres,strogatz2003sync,blekhmansynchronization,yoder2004unrolling,ramirez2020secret}. While confined to his room in February 1665 with a ``slight indisposition,'' Huygens noticed that two of the pendulum clocks he had recently built were swinging in perfect time together. Suspecting that they must be interacting somehow, perhaps through vibrations in their common support, Huygens did a series of experiments to test the idea. In one experiment, he attached two clocks to a board suspended on the backs of two chairs (Fig.~\ref{fig1}) and noticed, to his amazement, that no matter how he started the clocks, within about thirty minutes their pendulums always settled into antiphase synchrony, repeatedly swinging toward each other and then apart.  

In the centuries since then, and especially in the past twenty years, Huygens's clocks have been revisited by many authors, using a variety of experimental set-ups, simplified models, analytical approximations, and computational techniques~\cite{ellicott1740acount,ellicott1740further, ellis1873sympathetic, korteweg1906horloges, bennett2002huygens,oud2006study,senator2006synchronization,dilao2009antiphase, czolczynski2009clustering,czolczynski2011two, czolczynski2011huygens, czolczynski2013synchronization,kapitaniak2012synchronization, jovanovic2012synchronization,ramirez2013synchronization, ramirez2014improved, ramirez2014further, ramirez2016sympathy,ramirez2016poincare, ramirez2020secret, oliveira2015huygens,willms2017huygens,wiesenfeld2017huygens}. The subject has been reinvigorated by the rise of nonlinear science. New tools  have made it possible to simulate and solve the equations of motion for these coupled nonlinear  oscillators, and geometric ideas have illuminated many aspects of the system's dynamics and bifurcations.  

\begin{figure}
\centering
\includegraphics[width=1.8in]{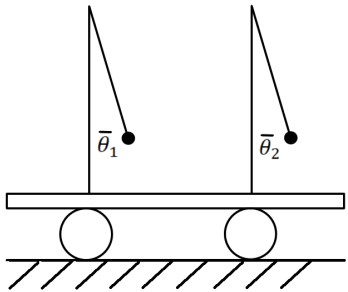}
\caption{\label{fig2} In-phase synchronization of two metronomes. The metronomes are drawn schematically, emphasizing the weight typically hidden inside the case.}
\end{figure}

As further motivation for studying this class of problems, consider the curious behavior of coupled metronomes~\cite{pantaleone2002synchronization,kuznetsov2007synchronization,ulrichs2009synchronization,wu2012anti}. Videos of self-synchronizing metronomes have attracted millions of views on YouTube~\cite{bahraminasab2007synchronisation,32metronomes,mythbusters2014nsync}. In these experiments, following the work of Pantaleone~\cite{ pantaleone2002synchronization}, anywhere from two to 32 metronomes are placed on a light platform that is free to move sideways, typically by rolling on empty soda cans or other light cylinders (Fig.~\ref{fig2}). As with Huygens's clocks,  synchronization gradually occurs after several minutes. A striking difference from Huygens's clocks, however, is that the metronomes in the videos tend to synchronize in phase rather than in antiphase, and one naturally wonders why.

Although a great deal of progress has been made in understanding the synchronization of clocks and metronomes, many questions remain~\cite{ramirez2020secret}. One challenge is to model the individual clocks, keeping in mind that real pendulum clocks come in many shapes, sizes, and styles. There are monumental clocks~\cite{ramirez2016sympathy} which stand upright like grandfather clocks. There are smaller clocks that can be placed on a mantel or hung on a wall~\cite{czolczynski2011huygens, oliveira2015huygens}. And there are stripped-down, simplified clocks designed for laboratory experiments~\cite{bennett2002huygens, wiesenfeld2017huygens}. The same bewildering diversity is true of metronomes. 

A further challenge is to model the escapement mechanism that keeps the clock or metronome running. Depending on the level of realism one seeks, the escapement can be modeled with discontinuous impulses~\cite{bennett2002huygens}, step functions~\cite{oliveira2015huygens}, Gaussian-like functions~\cite{oud2006study}, piecewise linear functions~\cite{ramirez2016sympathy} or a nonlinear damping term of the sort seen in van der Pol oscillators~\cite{pantaleone2002synchronization}. 

In addition to modeling the individual clocks, there is also the challenge of modeling the coupling between them. Clocks can interact by transmitting sound pulses through a wall on which they are both hanging~\cite{oliveira2015huygens}, or by shaking their common support or the beam from which they are suspended~\cite{ramirez2016sympathy, ramirez2020secret}. Metronomes interact by jiggling their shared platform slightly from side to side every time they swing, thereby transmitting phase information to each other~\cite{pantaleone2002synchronization}.

Along with these variations in experimental conditions, there is a correspondingly large variety of analytical methods that can be used to study such systems. Early researchers used linear techniques such as normal modes~\cite{korteweg1906horloges}. Others have applied qualitative geometric methods, such as Poincar\'{e} maps~\cite{bennett2002huygens} and equivariant bifurcation theory~\cite{willms2017huygens}. Still others have used perturbation methods, such as Poincar\'{e}’s method~\cite{ramirez2016poincare} or the method of multiple scales~\cite{pantaleone2002synchronization}. When the goal is to make quantitative predictions, many researchers have relied on numerical integration of the governing ordinary differential equations for a simplified three-degree-of-freedom model, or for further realism, Ramirez and colleagues have used finite element models of the coupling bar through which the clocks interact~\cite{ramirez2014improved}. 

Finally, along with all these methodological choices, there is a zoo of phenomena to be analyzed. Besides in-phase and antiphase synchronization, pendulum clocks sometimes exhibit quasiperiodicity as well as other forms of  unsynchronized behavior. In some circumstances, one of the clocks may stop ticking altogether. This  phenomenon~\cite{ellicott1740acount, ellicott1740further, bennett2002huygens}, now called ``beating death,'' occurs if one of the pendulums swings at such low amplitude that it fails to engage its escapement mechanism.

In this paper we focus on one of these many questions. Our goal is to clarify what causes clocks or metronomes to synchronize in phase or in antiphase. Our strategy is to recast the governing equations for such systems into a form amenable to perturbation theory, by rescaling them so that they become a weakly perturbed pair of harmonic  oscillators. Then the method of multiple scales yields the evolution equations for the oscillators' slowly-varying amplitudes and phases. These evolution equations   determine whether the system will ultimately synchronize, and if so, in which mode.  The perturbation methods we use are well known, but they have rarely been applied in this setting. 

Our work is closest in spirit to that of Pantaleone~\cite{pantaleone2002synchronization}. He used the method of multiple scales in a similar fashion to analyze the dynamics of coupled metronomes. But whereas he assumed a van der Pol approximation for the escapement mechanism, we were curious to see what qualitative differences might arise from a more realistic impulsive model of the escapement. We have also assumed a more general model of the coupling between the oscillators. Pantaleone~\cite{pantaleone2002synchronization} assumed that the metronomes rest on a platform that is neither damped nor subject to any restoring forces. We allow for both. By doing so, we find results consistent with some transition scenarios reported experimentally~\cite{wu2012anti}.  In particular, we find a sequence of transitions as the damping of the platform is increased. At low enough damping, and assuming the pendulums' amplitudes are small but not too small, the system tends to synchronize in phase (as seen in many  experiments on coupled metronomes). At intermediate damping, both in-phase and antiphase synchronization become stable, each with its own basin of attraction. And at sufficiently high damping, only antiphase synchronization is stable. These results agree with those found in the metronome experiments of Wu et al.~\cite{wu2012anti}

Other notable features of our approach are: (1) the attention given to modeling the escapement mechanism, (2) a small-angle approximation expanded past the linear term, and (3) the  scaling of the model equations that disentangles different physical effects through a two- and a three-timescale asymptotic analysis. While each of these ingredients can be found in the literature, this is the first time, to the best of our knowledge, that all three have been considered simultaneously. 

This article is organized as follows. In Section~\ref{s2}, we discuss the mechanics of pendulum clocks and metronomes, with particular emphasis on the modeling of the escapement mechanism. Section~\ref{s3} develops the rest of the model. In Section~\ref{s4} we identify the timescales over which the various physical effects take place. These considerations guide our choices of the relevant dimensionless parameters and variables. We call the resulting dimensionless system of equations the \emph{scaled system}. In Section~\ref{asymp} we carry out a two-timescale asymptotic analysis on the scaled system to obtain the \emph{slow flow} (often referred to in the perturbation theory literature as the averaged system). This slow flow governs the long-term dynamics. It has one fixed point that corresponds to in-phase synchronization, and another that corresponds to antiphase synchronization. In Section~\ref{s7}, we explore how the damping of the platform affects the stability of these synchronized states. The results are presented in a set of bifurcation diagrams, with explicit formulas given for the  bifurcation curves appearing in the diagrams. To simplify the analysis further, in Section~\ref{s8} we assume that the influence of the clocks or metronomes on the motion of the platform is extremely weak. This assumption produces an extra separation of timescales, and with it, some extra insight into  the underlying physics. The paper concludes in Section~\ref{s10} with a brief discussion.

\section{The Escapement Mechanism}
\label{s2}

We begin by describing the mechanics of the escapement mechanism~\cite{kapitaniak2012synchronization,
lepschy1992feedback,
moon2006coexisting,roup2003limit,
rowlings1944science}. This is the mechanism that provides the source of energy for  pendulum clocks and metronomes. 

\begin{figure}
\includegraphics[width=3in]{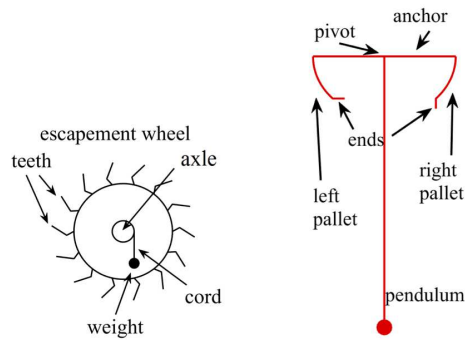}
\caption{\label{fig3} Components of our model clock.}
\end{figure}

The left panel of Fig.~\ref{fig3} shows the components of a so-called deadbeat anchor escapement in clocks (the escapements for metronomes work similarly, except their energy source is a spring that unwinds instead of a weight that descends).
The main components are the axle, the escapement wheel, and the weight. The escapement wheel is a gear with teeth. The axle extends in the direction perpendicular to the page and goes through the center of the escapement wheel. The escapement wheel and the axle rotate together. The weight provides energy to the system; it hangs from a cord wound around the axle and as it descends, it applies a torque to the axle to turn the axle-escapement wheel system in the clockwise direction. 

\begin{figure}
\centering
\includegraphics[width=2.8in]{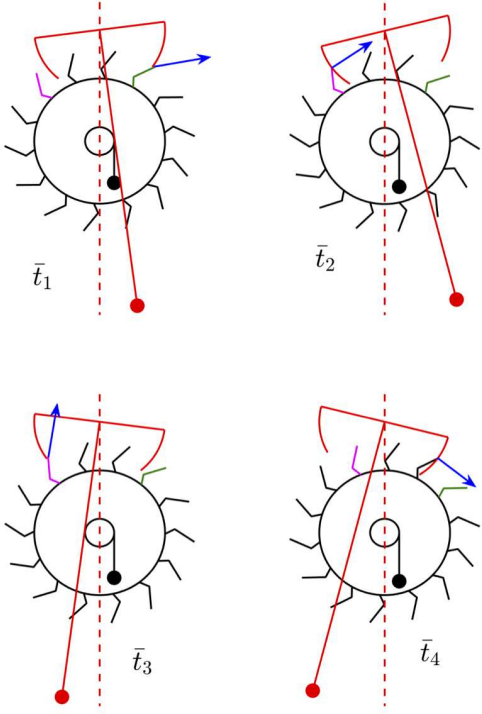} 
\caption{\label{fig5} Snapshots of the deadbeat anchor escapement at different points in its cycle.}
\end{figure}

The right panel of Fig.~\ref{fig3} shows a pendulum rigidly attached to an anchor. The sides of the anchor are known as pallets. The pendulum, anchor, and pallets all oscillate together about their common pivot, as shown in Fig.~\ref{fig5}, which in turn causes the teeth of the wheel to interact with the pallets. Whenever a tooth strikes a pallet, it does so without recoil; this is where the ``dead'' in ``deadbeat'' comes from. Moreover, a tooth in contact with a pallet slides along the pallet face \emph{without} applying torque to the system.  Torque is applied only when a tooth reaches the \emph{end} of a pallet. Note that the right and left ends of the pallets are differently shaped, as shown in  the right panel of Fig.~\ref{fig3}; this shape difference is crucial to obtain the desired clock dynamics (but, for visual clarity, those shape differences are suppressed in Fig.~\ref{fig5}).

To see how energy is transferred from the escapement to the pendulum, consider four key moments in a swing cycle (Fig.~\ref{fig5}). At time $\bar{t}_1$, the pendulum is swinging counterclockwise, and the green tooth (located near 1 o'clock on the escapement wheel) is contacting the end of the right pallet, thereby applying a force on it \emph{perpendicular} to the pallet's end (this is where the end shape of the pallet matters). Because the force points in the direction of the blue arrow shown in Fig.~\ref{fig5}, the anchor-pendulum system experiences an impulse that increases its kinetic energy.  

Once the green tooth is no longer in contact with the right pallet, the escapement wheel accelerates clockwise due to the torque caused by the weight. Then the escapement wheel stops abruptly when the pink tooth (located near 11 o'clock on the wheel) meets the left pallet. Meanwhile, the anchor-pendulum system continues turning counterclockwise. At time $\bar{t}_2$ in Fig.~\ref{fig5}, the pendulum makes its largest angle with the vertical.  While the pink tooth is in contact with the pallet face, the tooth applies a force that points toward the pivot of the anchor-pendulum system (because the pallet face is a circular arc at a constant radial distance from the pivot). Hence this force does \emph{not} apply any torque to the anchor-pendulum system with respect to the pivot, and so the dynamics of the anchor-pendulum system is not affected when the tooth is in contact with the pallet face. 

Similar events occur in the next half of the cycle, with times $\bar{t}_3$ and $\bar{t}_4$ playing the parts of $\bar{t}_1$ and $\bar{t}_2$. Energy is pumped into the pendulum at time $\bar{t}_3$, and only then. 

The self-sustained oscillations of the pendulum continue until the cord that holds the weight is no longer wound around the axle of the escapement wheel. The periodic input of energy that the anchor-pendulum system receives from the escapement wheel-weight system makes up for the energy lost due to damping. 

\section{Model of two coupled clocks}
\label{s3}

\begin{figure}
\centering
\includegraphics[width=3in]{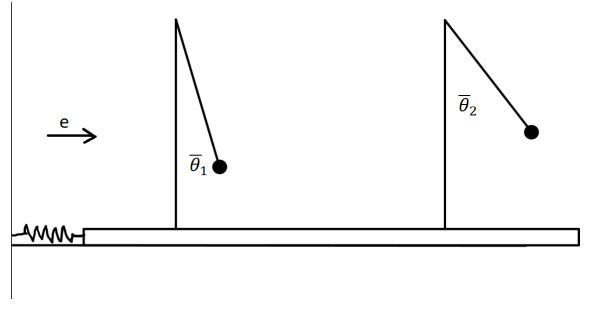}
\caption{\label{setup} Cartoon of the two clocks/metronomes and the platform.}
\end{figure}

A cartoon of our model is shown in Fig.~\ref{setup}. Both of the pendulum angles  $\bar{\theta}_1$ and $\bar{\theta}_2$ are functions of time $\bar{t}$. We use primes to denote derivatives with respect to $\bar{t}$. (The overbars denote dimensional quantities; they will disappear soon, after we nondimensionalize the system.) 
 
To keep track of the motion of the platform or support, we select an arbitrary point on it. The position of this point is denoted by $\bar{x}\,{\bf e}$, where ${\bf e}$ is the constant unit dimensionless vector that points to the right, as illustrated in Fig.~\ref{setup}. Note that for simplicity we are regarding the platform as having only one degree of freedom. 
 
 To model the action of the escapement on pendulum $i$, where $i=1,2$, we assume there is a constant impulse $\bar{J}$ and a critical angle $\bar{\theta}_c$ such that pendulum $i$ receives a positive impulse $\bar{J}$ whenever it reaches its critical angle while swinging to the right, i.e., whenever  $\bar{\theta}_i = \bar{\theta}_c$ and  $\bar{\theta}_i^{'}>0$. Such an impulse occurs at time $\bar{t}_1$ in Fig.~\ref{fig5}.  Similarly, a negative impulse $-\bar{J}$ is received whenever $\bar{\theta}_i = -\bar{\theta}_c$ and $\bar{\theta}_i^{'}<0$ (as at time $\bar{t}_3$ in Fig.~\ref{fig5}). Let $\{\bar{T}_{i r}\}$ be the set of times when pendulum $i$ receives a positive impulse, and let  $\{\bar{T}_{i \ell}\}$ be the set of times when it receives a negative impulse.
We define
\begin{eqnarray}
\nonumber
\bar{f}_1(\bar{t}) = \sum_{\bar{t}_\star\in \bar{T}_{1 r}} \bar{J} \delta(\bar{t}-\bar{t}_\star) - \sum_{\bar{t}_\star\in \bar{T}_{1 \ell}} \bar{J} \delta(\bar{t}-\bar{t}_\star) 
\end{eqnarray}
and
\begin{eqnarray}
\nonumber
\bar{f}_2(\bar{t}) = \sum_{\bar{t}_\star\in \bar{T}_{2 r}} \bar{J} \delta(\bar{t}-\bar{t}_\star) - \sum_{\bar{t}_\star\in \bar{T}_{2 \ell}} \bar{J} \delta(\bar{t}-\bar{t}_\star),
\end{eqnarray}
where $\delta$ is the delta function. 

This model of the escapement mechanism is idealized. It is the simplest model that, in our opinion, contains the main physics relevant to our studies. More realistic models of the escapement mechanism assume that the input of energy to the pendulum is not applied instantaneously as an impulse, but rather as a force applied over a short but nonzero period of time. Such forces have been modeled with Gaussian-like~\cite{oud2006study} or piecewise linear~\cite{ramirez2016sympathy} functions. Although they differ in detail, these alternative escapement models are not qualitatively different from ours. As long as the period of time over which the force acts is much shorter than the time of an oscillation, these short-acting forces can be approximated by delta functions. 

 Having modeled the escapement forces, we turn now to the forces on the coupling platform. We assume that it is subjected to a linear restoring force ${\bar {\bf F}}_r$ that follows  Hooke's law,  ${\bar {\bf F}}_r = -\bar{\kappa}\, \bar{x}\, {\bf e}$. Here, the origin of $\bar{x}$ has been implicitly chosen such that the restoring force is equal to ${\bf 0}$ when $\bar{x}=0$. The platform is also assumed to be subjected to a linear damping force ${\bar {\bf F}}_d = -\bar{\mu}  \bar{x}^{'} {\bf e}$.
 
The remaining parameters in the equations of motion are as follows: $m$ is the mass of each pendulum; $M$ is the combined mass of both metronomes or clocks, including their pendulums, and the platform; $L$ is the length of each pendulum, namely the distance from the pivot to the center of mass of the pendulum; $\bar{\nu}$ is a damping constant (due to the pendulum motion); and $g$ is the acceleration due to gravity. For simplicity, we neglect the mass of the escapement wheels. Then Newton's second law yields the following system, which we refer to as the \emph{governing equations}:
\begin{align*}
m L \bar{\theta}_1^{''} = - m g \sin \bar{\theta}_1 - \bar{\nu} L  \bar{\theta}_1^{'} - m \bar{x}^{''} \cos \bar{\theta}_1  +\bar{f}_1 \\
m L \bar{\theta}_2^{''} = - m g \sin \bar{\theta}_2 - \bar{\nu} L \bar{\theta}_2^{'} - m \bar{x}^{''} \cos \bar{\theta}_2 +\bar{f}_2 \\
M \bar{x}^{''} = 
- m L \left( \bar{\theta}_1^{''} \cos \bar{\theta}_1  - \bar{\theta}_1'^2 \sin \bar{\theta}_1 +  \bar{\theta}_2^{''} \cos \bar{\theta}_2 - \bar{\theta}_2'^2 \sin \bar{\theta}_2 \right)\\ -\bar{\kappa}\bar{x} -\bar{\mu} \bar{x}^{'}.
\end{align*}
The first and second equations are obtained by taking torques about the pivots of the corresponding pendulum and  dividing by $L$. The third equation expresses horizontal force balance for the platform.  

The governing equations apply to both clocks and metronomes.  For example, in the original Huygens set-up shown schematically in Fig.~\ref{fig1}, the restoring force on the support is zero but the damping force on it is not. Thus, if the damping force is assumed to be proportional to the support velocity, the dynamics of the system in Fig.~\ref{fig1} is mimicked by the cartoon in Fig.~\ref{setup} and by the equations above with $\bar{\kappa} = 0$ and $\bar{\mu} \neq 0$. Similarly, Fig.~\ref{fig2} is a simplified model of the metronome experiments. If the rollers underneath the coupling platform are assumed to be massless and do not slip, the dynamics of the system in Fig.~\ref{fig2} are given by the equations above with $\bar{\mu} = \bar{\kappa} = 0$. In what follows, we will analyze the governing equations for general values of $\bar{\mu}$ and $\bar{\kappa}$, but with special attention to cases where one or both of these are zero.


\section{Characteristic scales and nondimensionalization}
\label{s4}

Since synchronization takes place after hundreds of swings, the relevant physics occurs at two different timescales.  
We introduce a small parameter $\epsilon \ll 1$  that encodes the separation of these timescales. We will choose the rest of the dimensionless parameters and variables so that the different physical effects take place on either the timescale of a single swing of a pendulum, or a much longer timescale given by $1/\epsilon$ times the pendulum's period. 

Specifically, we scale the variables and parameters as follows. The natural choice for the dimensionless time $t$ is $$t = \bar{t} \sqrt{g/L} $$ so that the periods of the pendulums are $O(1)$ in $t$. An $O(1)$ phase adjustment of the pendulums, due to inertial forcing from the motion of the platform, occurs in long times of $O(M/m)$ in $t$; thus, we want $M/m = O(1/\epsilon),$ or equivalently, the mass ratio $m/M = O(\epsilon).$ This choice leads us to introduce a dimensionless parameter $$ \epsilon b = \frac{m}{M},$$ where  $b$ is assumed to be $O(1)$. Physically, $b$ quantifies how strongly the  pendulums' motion affects the platform's motion. Consequently, $b$ also controls how much one pendulum couples to the other, mediated by the driving they each impart to the platform. (All of this will become clearer after we nondimensionalize the governing equations; see Eq.~(\ref{equ}) below.)

To scale the angle variables, note first that the $\bar{\theta}_i$ are of the order $\bar{\theta}_c$, the critical angle at which the escapement engages. To make the nonlinear equations of motion as tractable as possible, we want to use a small-angle approximation, but we also want to retain the leading effects of nonlinear terms. With these ideas in mind, note that $\sin \bar{\theta}_i \approx \bar{\theta}_i + O(\bar{\theta}_i^3)$, so the leading nonlinear effects take place in times of $O(1/\bar{\theta}_i^{2})$ in $t$. Thus, we want $1/\bar{\theta}_c^2 = O(1/\epsilon)$, which motivates the following scaling: 
$$\theta_c = \bar{\theta}_c/\sqrt{\epsilon r,}\;\;\;\;\;\;\theta_i = \bar{\theta}_i/\sqrt{\epsilon r},$$ 
where the dimensionless parameters $r$ and $\theta_c$ are  $O(1)$.

To scale the remaining quantities in the model, we estimate that $\bar{x} = O(L \bar{\theta}_i m/M)$. Since $\bar{\theta}_i = O(\sqrt{\epsilon r})$ and $m/M = O(\epsilon)$, we introduce $$x = \bar{x} / (L \epsilon \sqrt{\epsilon r}),$$ so that $x$ is $O(1)$. The damping due to friction in the pendulums takes place in times of $O\left((m/\bar{\nu})\sqrt{g/L}\right)$ in $t$. Since we want this quantity to be $O(1/\epsilon)$, we introduce the $O(1)$ dimensionless parameter $$\nu = (\bar{\nu}/m\epsilon) \sqrt{L/g}.$$ 

The impulse $\bar{J}$ causes an increase in the amplitude of oscillations of $O\left(\bar{J}/(m\sqrt{gLr\epsilon})\right)$ in the dimensionless variables $\theta_i$. We want this quantity to be $O(\epsilon)$ so the cumulative effects of the impulses take  place in times of $O(1/\epsilon)$ in $t$. Therefore we define $$J = \epsilon^{-3/2} \bar{J}/(m \sqrt{gLr}),$$ and assume it to be $O(1)$. 

The restoring force and damping force on the platform should affect the dynamics of the platform in times of $O(1)$. This leads to the following dimensionless stiffness and damping parameters for the platform: $$\kappa = \frac{L\bar{\kappa}}{M g}, \qquad  \mu = \frac{\bar{\mu}}{M} \sqrt{\frac{L}{g}}.$$

Next, we nondimensionalize the governing equations. Let $\{T_{ir}\}$ and $\{T_{il}\}$ be the set of dimensionless times $t$ when pendulum $i$ receives a positive or negative impulse, respectively. We define 
\begin{eqnarray}
\label{f1}
f_1(t) = \sum_{t_\star\in T_{1r}}  \delta(t-t_\star) - \sum_{t_\star\in T_{1\ell}}  \delta(t-t_\star)
\end{eqnarray}
and
\begin{eqnarray}
\label{f2}
f_2(t) = \sum_{t_\star\in T_{2r}}  \delta(t-t_\star) - \sum_{t_\star\in T_{2\ell}}  \delta(t-t_\star).
\end{eqnarray}
Then, neglecting terms of  $O(\epsilon^2)$ in the first two governing equations and terms of  $O(\epsilon)$ in the third equation, we find that the governing equations become
\begin{align}
\nonumber
 \ddot{\theta}_1 + \theta_1 &= \epsilon \frac{r}{6} \theta_1^3 - \epsilon \nu \dot{\theta_1} + \epsilon J f_1 -\epsilon \ddot{x}\\ 
\label{equ}
\ddot{\theta}_2 + \theta_2 &= \epsilon \frac{r}{6} \theta_2^3 - \epsilon \nu \dot{\theta}_2 +\epsilon J f_2 - \epsilon \ddot{x}\\ 
\nonumber
\ddot{x} + \mu \dot{x} + \kappa x &= - b\left(\ddot{\theta}_1 + \ddot{\theta}_2 \right),
\end{align}
where dots denote derivatives with respect to $t$. We refer to Eq.~(\ref{equ}) as the \emph{scaled system}. 

Notice that our choice of scaling has converted the first two governing equations into undamped linear oscillators perturbed by various forces of size $O(\epsilon)$. Although these forces are small, their effects accumulate on a long timescale of $O(1/\epsilon)$. The energy source that drives this slow evolution is the train of small impulses coming from the escapements (the $\epsilon J f$ terms). Meanwhile, the parts of the system interact through two-way coupling: the pendulums are inertially forced by the platform's accelerated motion (the $\epsilon \ddot{x}$ terms), while the pendulums act back on the platform through the $b \ddot{\theta}$ terms. 

In deriving this scaled system, we kept only the dominant terms, as is customary in perturbation theory. The terms of $O(\epsilon^2)$ neglected in the first two equations can be proven not to affect the existence or stability of the in-phase or antiphase synchronized states, or of any other periodic solutions, as long as those periodic solutions satisfy certain genericity conditions; this fact follows from a basic theorem of averaging theory~\cite{guckenheimer1983nonlinear}. Admittedly, the $O(\epsilon^2)$ terms can make tiny quantitative changes to the solutions, but these are asymptotically negligible for $\epsilon \ll 1$. For the same reason, it suffices to consider only the $O(1)$ terms in the third equation. 

However, the leading perturbations of size $O(\epsilon)$ in the first two equations must be retained. As is well known from the theory of weakly nonlinear oscillators~\cite{guckenheimer1983nonlinear,strogatz1994nonlinear, bender1999advanced, holmes1995introduction}, these small perturbations play a decisive role in determining the system's long-term qualitative behavior.  

The structure of Eq.~(\ref{equ}) also clarifies what the  parameter $\epsilon$ represents physically. A clue to its meaning is that the only terms with a pure coefficient of $\epsilon$ and no other prefactors are the $\epsilon \ddot{x}$ terms in the first two equations in (\ref{equ}). Then, by tracing this clue back to the original governing equations, one can check that $\epsilon$ is a dimensionless quotient of two characteristic accelerations: the typical accelerations $\bar{x}^{''}$ of the platform and the (much larger) typical  accelerations $L \bar{\theta}^{''}$ of the pendulums. 

The cubic terms multiplied by $ \epsilon r$ in Eq.~(\ref{equ}) are not usually considered, but turn out to be important. As we will see in subsequent sections, in the parameter regimes of interest the size of the cubic coefficient $r$ determines whether the pendulums will synchronize in phase or in antiphase, or whether both of those synchronized states are locally stable. From a physical standpoint, increasing $r$ corresponds to increasing the critical angle $\bar{\theta}_c$ and the impulse $\bar{J}$ in the same proportion while keeping all the other dimensional parameters constant. Because of this close connection between $r$ and $\bar{J}$, we will sometimes find it helpful to interpret $r$ as a dimensionless measure of the impulsive forcing strength, even though $J$ plays this role more directly.


\section{Perturbation Analysis}
\label{asymp}

To make the analysis as clear as possible, we begin with the simplest case: $\mu=0$ and $\kappa=0$.  (The procedure for analyzing the more general case when these two parameters are nonzero is similar to that presented below. But the resulting equations are complicated enough to obscure the benefits of the method.  The messy asymptotic equations for $\mu \neq 0$ and $\kappa \neq 0$ are relegated to Appendix~\ref{s5}.)

As discussed in previous sections, we are assuming the mass ratio $m/M$ is of the same order as a small parameter $\epsilon \ll 1$. With suitable scaling of the other physical parameters, the dynamics then take place on two timescales, one of $O(1)$ and the other of $O(1/\epsilon)$ in $t$. Specifically, the pendulum angles $\theta_{1}$, $\theta_{2}$ and the dimensionless location $x$ of the system's center of mass are oscillatory variables with periods of $O(1)$ in $t$, but their amplitudes and phases change by $O(1)$ on timescales of $O(1/\epsilon)$ in $t$. Thus, we make the following ansatz:
\begin{equation}
\label{an}
\begin{aligned}
\theta_{i}(t) & \sim\theta_{i0}(t,\tau) + \epsilon \theta_{i1}(t,\tau) + \cdots,\quad i = 1, 2 \\
x(t) & \sim x_0(t, \tau) + \epsilon x_1(t,\tau) + \cdots
\end{aligned}
\end{equation}
where $\sim$ means asymptotic approximation in the parameter regime $\epsilon\ll 1$, and $$\tau=\epsilon t$$ is a slow time variable. Each $\theta_{ij}$ and $x_i$ are functions of $t$ and $\tau$, and these functions are periodic in their first argument $t$.   

We carry out a standard two-timescale analysis\cite{guckenheimer1983nonlinear, strogatz1994nonlinear, holmes1995introduction,bender1999advanced}. Namely, we plug the ansatz~(\ref{an}) into the system of equations~(\ref{equ}), then replace $d/dt$ by
\begin{eqnarray}
 \frac{\partial}{\partial t} + \epsilon \frac{\partial }{\partial \tau}
\end{eqnarray}
in that system (this substitution follows from the form of the ansatz~(\ref{an})), and finally collect terms having like powers of $\epsilon$. This perturbative method, often called two-timing, is a special case of the method of multiple scales\cite{strogatz1994nonlinear,holmes1995introduction,bender1999advanced}. It can be rigorously justified by averaging theory\cite{guckenheimer1983nonlinear}. 

From the terms that contain the power $\epsilon^0$ in the expansion of (\ref{equ}), we obtain 
\begin{equation*}
\begin{aligned}
\frac{\partial^2\theta_{10}}{\partial t^2}  + \theta_{10} &=0\\
\frac{\partial^2\theta_{20}}{\partial t^2}  + \theta_{20} &=0\\
\frac{\partial^2 x_0}{\partial t^2}  +b \left( \frac{\partial^2 \theta_{10}}{\partial t^2}  + \frac{\partial^2\theta_{20}}{\partial t^2}  \right) &=0.
\end{aligned}
\end{equation*}
The general solution of these equations (recalling that $x_0$ is periodic in $t$) is
\begin{eqnarray}
\label{as10}
\theta_{10}(t,\tau) &= A_1(\tau) \, \sin(t+\varphi_1(\tau)) \\
\label{as20}
\theta_{20}(t,\tau) &= A_2(\tau) \, \sin(t+\varphi_2(\tau))
\end{eqnarray}
and 
\begin{align*}
x_0=-b(\theta_{10}+\theta_{20}).
\end{align*}

As usual, differential equations for the evolution of the slow variables $A_1,A_2,\varphi_1, \varphi_2$ will be obtained at the next order of $\epsilon$. But before we proceed to that order, we need to deal with an unusual feature of our model system (\ref{equ}): it contains delta-function forcing terms due to the repeated impulses provided by the escapement mechanism. Now that we have an asymptotic approximation for the fast oscillations of the pendulums, we can find the times when the escapement acts; by solving for these times and inserting them into the delta functions, we get the following asymptotic approximations for the impulsive forcing terms $f_1$ and $f_2$ in Eqs.~(\ref{f1}) and~(\ref{f2}):  
\begin{eqnarray}
\label{af12}
f_1(t) \sim f_{10}(t,\tau) \mbox{ and }
f_2(t) \sim f_{20}(t,\tau),
\end{eqnarray}
where
\begin{eqnarray}
\nonumber
f_{10}(t,\tau) = \sum_{n \in {\mathbb Z}}  \delta \left( t - \arcsin \left( \frac{\theta_c}{A_1(\tau)} \right) + \varphi_1 (\tau) + 2 n \pi \right) \\ \nonumber
-  \sum_{n \in {\mathbb Z}}  \delta \left( t - \arcsin \left( \frac{\theta_c}{A_1(\tau)} \right) + \varphi_1 (\tau) + (2 n + 1) \pi \right) 
\end{eqnarray}
and
\begin{eqnarray}
\nonumber
f_{20}(t,\tau) = \sum_{n \in {\mathbb Z}}  \delta \left( t - \arcsin \left( \frac{\theta_c}{A_2(\tau)} \right) + \varphi_2 (\tau) + 2 n \pi \right) \\ \nonumber
-  \sum_{n \in {\mathbb Z}}  \delta \left( t - \arcsin \left( \frac{\theta_c}{A_2(\tau)} \right) + \varphi_2 (\tau) + (2 n + 1) \pi \right).
\end{eqnarray}

Now proceeding to the $O(\epsilon^1$) terms in the expansion of the system~(\ref{equ}), we find that its first two equations give
\begin{equation}
\begin{aligned}
\frac{\partial^2\theta_{11}}{\partial t^2} + \theta_{11} &= \frac{r}{6} \theta_{10}^3 - \nu \frac{\partial \theta_{10}}{\partial t} + J f_{10} - \frac{\partial^2 x_0}{\partial t^2} - 2 \frac{\partial^2\theta_{10}}{\partial t \partial \tau} 
\\ 
\frac{\partial^2\theta_{21}}{\partial t^2} + \theta_{21} &= \frac{r}{6} \theta_{20}^3 - \nu \frac{\partial \theta_{20}}{\partial t} + J f_{20} - \frac{\partial^2 x_0}{\partial t^2} - 2 \frac{\partial^2\theta_{20}}{\partial t \partial \tau}. 
\end{aligned}
\label{eq-order-eps}
\end{equation}

Next, to derive the slow flow equations for $A_1,A_2,\varphi_1, \varphi_2$, recall an elementary fact from the solvability theory of differential equations: {\it Let $h(t)$ be a $2\pi$-periodic function of $t$. Let $\varphi$ be any fixed real number. The equation $\ddot{\theta} + \theta = h$ has a $2\pi$-periodic solution $\theta$ if and only if $\int_{0}^{2\pi} h(t) \sin(t+\varphi) dt = 0$ and $\int_{0}^{2\pi} h(t) \cos(t+\varphi) dt = 0$.} This fact is usually stated with $\varphi=0$, but in our analysis it will be convenient to use $\varphi=\varphi_1$ and $\varphi=\varphi_2$.

The next step is to go back to  Eq.~(\ref{eq-order-eps}), recall that $\theta_{ij}$ are $2\pi$-periodic in $t$, and use the fact stated in the last paragraph to conclude that  
\begin{eqnarray}
\int_{0}^{2\pi} \left( \frac{r}{6} \theta_{10}^3 - \nu \frac{\partial \theta_{10}}{\partial t} + J f_{10} - \frac{\partial^2 x_0}{\partial t^2} - 2 \frac{\partial^2\theta_{10}}{\partial t \partial \tau} \right) \nonumber
\\ \nonumber \times \sin(t+\varphi_1)\,dt=0,
\end{eqnarray}
\begin{eqnarray}
\nonumber
\int_{0}^{2\pi} \left( \frac{r}{6} \theta_{10}^3 - \nu \frac{\partial \theta_{10}}{\partial t} + J f_{10} - \frac{\partial^2 x_0}{\partial t^2} - 2 \frac{\partial^2\theta_{10}}{\partial t \partial \tau} \right) \\ \times \cos(t+\varphi_1) \,dt=0,
\nonumber
\end{eqnarray}
\begin{eqnarray}
\nonumber
\int_{0}^{2\pi} \left( \frac{r}{6} \theta_{20}^3 - \nu \frac{\partial \theta_{20}}{\partial t} + J f_{20} - \frac{\partial^2 x_0}{\partial t^2} - 2 \frac{\partial^2\theta_{20}}{\partial t \partial \tau} \right) \\ \times \sin(t+\varphi_2)\,dt=0
\nonumber
\end{eqnarray}
and
\begin{eqnarray}
\nonumber
\int_{0}^{2\pi} \left( \frac{r}{6} \theta_{20}^3 - \nu \frac{\partial \theta_{20}}{\partial t} + J f_{20} - \frac{\partial^2 x_0}{\partial t^2} - 2 \frac{\partial^2\theta_{20}}{\partial t \partial \tau} \right) \\ \times \cos(t+\varphi_2)\,dt=0.
\nonumber
\end{eqnarray}

By computing these four integrals (and omitting the algebraic details, which are long but straightforward), we obtain the following slow flow equations:
\begin{eqnarray}
\label{as1}
\frac{d A_1 }{d \tau} = -\frac{\nu}{2} A_1 + \sqrt{1-\frac{\theta_c^2}{A_1^2}} \, \frac{J}{\pi} + \frac{b}{2} A_2\sin(\varphi_1-\varphi_2)  \quad\\ 
\label{as2}
A_1 \frac{d \varphi_1}{d \tau} = \frac{b}{2} A_1 - \frac{\theta_c}{A_1} \, \frac{J}{\pi} + \frac{b}{2} A_2\cos(\varphi_1-\varphi_2)  -\frac{r}{16} A_1^3 \quad \\
\label{as3}
\frac{d A_2 }{d \tau} = -\frac{\nu}{2} A_2 + \sqrt{1-\frac{\theta_c^2}{A_2^2}} \, \frac{J}{\pi} + \frac{b}{2} A_1\sin(\varphi_2-\varphi_1)  \quad\\ 
\label{as4}
A_2 \frac{d \varphi_2}{d \tau} = \frac{b}{2} A_2 - \frac{\theta_c}{A_2} \, \frac{J}{\pi} + \frac{b}{2} A_1\cos(\varphi_2-\varphi_1)  -\frac{r}{16} A_2^3. \quad
\end{eqnarray}
This system holds for $A_1(\tau) > \theta_c$ and $A_2(\tau) > \theta_c$, meaning that the pendulums' swings are large enough to engage the escapement mechanism at all times.

Since our goal is to identify whether the system evolves to antiphase or in-phase synchronization or no synchronization at all, the variable of interest to us is the phase difference $$\psi = \varphi_1 - \varphi_2.$$ Dividing Eq.~(\ref{as2}) by $A_1$, dividing  Eq.~(\ref{as4}) by $A_2$, and subtracting the results, we obtain 
\begin{eqnarray}
\label{psi-simp}
\frac{d\psi}{d\tau} = \theta_c \, \frac{J}{\pi} \left(A_2^{-2}-A_1^{-2}\right) +\frac{r}{16}  \left(A_2^{2}-A_1^{2}\right) + \\ + \frac{b}{2} \left(\frac{A_2}{A_1} - \frac{A_1}{A_2}\right) \cos \psi. 
\nonumber 
\end{eqnarray}
We rewrite Eqs.~(\ref{as1}) and~(\ref{as3}) in terms of $\psi$ as
\begin{eqnarray}
\label{as1a}
\frac{d A_1 }{d \tau} = -\frac{\nu}{2} A_1 + \sqrt{1-\frac{\theta_c^2}{A_1^2}}\, \frac{J}{\pi} + \frac{b}{2}A_2 \sin\psi  \\ 
\label{as3a}
\frac{d A_2 }{d \tau} = -\frac{\nu}{2} A_2 + \sqrt{1-\frac{\theta_c^2}{A_2^2}} \,\frac{J}{\pi} - \frac{b}{2}A_1 \sin\psi. 
\end{eqnarray}
Equations~(\ref{psi-simp}),~(\ref{as1a}) and~(\ref{as3a}) form the slow flow system for the special case $\mu=0, \kappa=0$, in which we neglect the damping and restoring forces on the platform. The more general version, where $\mu$ and $\kappa$ are allowed to be nonzero, is given in Appendix~\ref{s5} as Eqs.~(\ref{as1full}),~(\ref{as3full}), and~(\ref{psifull}). 

\subsection{Stability analysis of in-phase and antiphase synchronization}

 The system~(\ref{psi-simp}),~(\ref{as1a}) and~(\ref{as3a}) has four obvious fixed points. Two of them are unstable for all values of the parameters, so we will ignore them from now on. The two fixed points that we will be concerned with are: (1) the in-phase fixed point $\psi=0$ with both pendulums swinging at a steady-state amplitude $A_1 = A_2 = A_s$; and (2) the antiphase fixed point $\psi = \pi$, again with $A_1 = A_2 = A_s$. In both cases, $A_s$ is given by 
 \begin{eqnarray}
 A_s = \sqrt{2}\,\frac{\theta_c}{\alpha} \sqrt{1+\sqrt{1-\alpha^2}},
 \nonumber
 \end{eqnarray}
where
$$\alpha = \frac{\pi\theta_c\nu}{J}.$$

It makes sense that this new dimensionless parameter $\alpha$ should enter this calculation. In physical terms, $\alpha$ is proportional to the  ratio between the damping force and the impulsive force on the pendulums. As one would expect, the balance between these two forces -- one which provides energy, and the other which dissipates it -- determines the long-term amplitudes of the pendulums' oscillations.

The eigenvalues of the Jacobian matrix associated with the system~(\ref{psi-simp}),~(\ref{as1a}), and~(\ref{as3a}) can be found explicitly at these synchronized states, and thereby provide information about their stability. We summarize the results here and direct the reader to Appendix~\ref{stab} for the derivation. Boundaries for stable in-phase and antiphase oscillations are given by the lines
\begin{equation}
\begin{aligned}
b_1(r) &= \frac{\nu \alpha}{1+\sqrt{1-\alpha^2}} - \frac{r \theta_c^2}{4 \alpha^2} \left(1+\sqrt{1-\alpha^2}\right), \\
b_2(r) &=- \frac{\nu \alpha}{1+\sqrt{1-\alpha^2}} + \frac{r \theta_c^2}{4 \alpha^2} \left(1+\sqrt{1-\alpha^2}\right).
\label{r-eqs}
\end{aligned}
\end{equation}
In-phase synchronization is unstable for $b<b_1(r)$ and locally stable for $b>b_1(r)$, whereas antiphase synchronization is unstable for $b<b_2(r)$ and locally stable for $b>b_2(r)$. 

In following sections, the value of $r$ where these two lines intersect is an important bifurcation point for the behavior of the system.  We call this point $r_{c}.$ As we will see below, $r_c$ marks the point above which antiphase synchronization is destabilized in favor of in-phase synchronization. The value of $r_c$ is given by
\begin{align}
r_{c}=\frac{4\nu \alpha^3}{\theta_{c}^2 \left(1+\sqrt{1-\alpha^2}\right)^2} .
\label{r_crit} 
\end{align}

\begin{figure}
\centering
\includegraphics[width=3.25in]{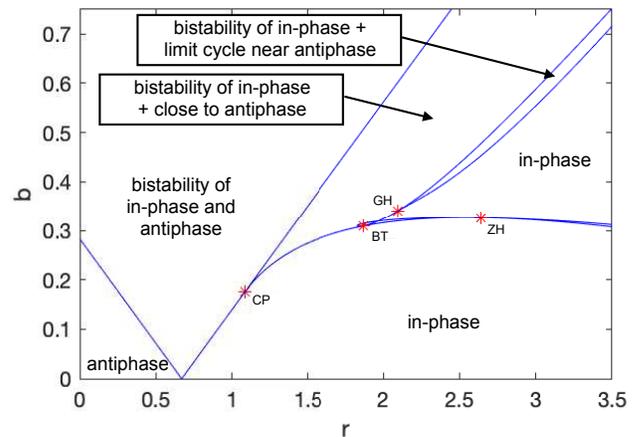} 
\caption{\label{r_vs_b} Bifurcation diagram in the parameters $b$ and $r$. The other parameters are fixed at the values $\theta_c=0.5$, $\nu =1$, $J=3$, $\mu=0 $ and $\kappa=0$. Recall that $b$ is proportional to the mass ratio $m/M$. In the scaled system Eq.~(\ref{equ}), $b$ appears as the strength of the back-coupling of the pendulums on the platform; it is proportional to the inertial force (due to the swinging of the pendulums) that drives the platform's motion. The cubic coefficient $r$ measures the strength of pendulums' frequency-dependence on amplitude, but it can be more usefully interpreted as a driving strength in its own right;  as $r$ increases, the impulsive forcing $\bar J$ on each pendulum increases, since $\bar J \propto J r^{1/2}$. The straight lines are graphs of the linear functions $b_1(r)$ and $b_2(r)$ given in Eq.~\eqref{r-eqs}.  These lines intersect the $r$-axis at the point $r_c$ given by Eq.~\eqref{r_crit}. The other curves were generated with the numerical bifurcation program {\sc Matcont} \cite{dhooge2008new}.} 
\end{figure}

Figure~\ref{r_vs_b} shows a bifurcation diagram in the parameters $b$ and $r$. The values of the other parameters are $\theta_c=0.5$, $\nu =1$, and $J=3$. The labels on the diagram indicate which states are locally stable in each part of parameter space. The straight lines correspond to the stability boundaries $b=b_1(r)$ and $b=b_2(r)$ in Eq.~(\ref{r-eqs}). The other curves in the diagram were generated with the numerical bifurcation program {\sc Matcont} \cite{dhooge2008new}. 

Fix $b$ to be small, say $b=0.1$. Then for small values of $r$, only antiphase oscillations are stable; for intermediate values of $r$, both forms of synchronization are stable; and for large values of $r$, only in-phase oscillations are stable. 

To interpret these results physically, recall that $r$ is a dimensionless measure of the pendulum's nonlinearity, which can become important when the  oscillations are small but not too small. Indeed, $r$ arose when we scaled the size of the critical angle at which the escapement engages and impulses are imparted. Our analysis shows that the nonlinear effects captured by $r$ are not negligible perturbations; on the contrary, they completely change the picture. We would not see a transition from antiphase to in-phase synchronization without them. Indeed, when $r=0$, Fig.~\ref{r_vs_b} shows that the antiphase state is always locally stable. To destabilize it, we need $r$ to be sufficiently large. 

We have also seen that $r$ reflects the  dependence of a pendulum's frequency on its amplitude, an effect that becomes increasingly important at large amplitudes.
In short, as the amplitudes increase, antiphase synchronization loses stability in favor of in-phase synchronization. As noted previously by Pantaleone~\cite{pantaleone2002synchronization} in the context of a different model, this finding may shed some light on why metronomes tend to synchronize in phase: they have a larger critical angle and typically swing at much larger amplitudes than the pendulums in pendulum clocks.

For larger fixed values of the coupling constant $b$, and for larger values of $r$, Fig.~\ref{r_vs_b} shows a more complicated scenario.  In particular, for $r$ increasing from small values, the stable fixed points corresponding to antiphase states branch into two stable equilibria with a phase difference $\psi$ that is nearly, but not exactly, equal to $\pi$; meanwhile, the exactly antiphase states lose stability.  For slightly larger values of $r$, the two stable, nearly antiphase oscillations bifurcate into two limit cycles in a supercritical Hopf bifurcation.  Finally, at even larger values of $r$, the limit cycles lose their stability and only in-phase synchrony is stable.  We were able to find the  nearly-antiphase states and the stable limit cycles in numerical simulations of the original nondimensional equations.

\section{Testing the model}
\label{s7}

There are many physical parameters we could vary to test the model, but we have chosen to focus on the platform damping $\bar{\mu}$, as it is one of the easiest parameters to adjust experimentally. Before delving into the predictions of our model, let us recall what has been seen in the lab.

\subsection{Experimental studies of platform damping}

Wu et al.\cite{wu2012anti} did an experiment with two metronomes placed on a platform that rolled on wheels of different radii: large, medium, and small. The wheels themselves rolled on surfaces that provided different amounts of friction: ``slippery'' glass board, coarse cloth, and five pieces of coarse cloth.  Wu et al. conducted 100 tests each for the three different wheel sizes and the three different  surfaces. They found that as  the rolling friction increased, the system went from having only stable in-phase synchronization at low friction, to both in-phase and antiphase synchronization at medium friction (with the outcome changing from test to test, depending on the initial conditions), to only antiphase synchronization at high friction.  

When the friction was lowest (under conditions with large wheels or when rolling on the glass surface), all 100 tests  synchronized in phase.  At the other extreme, when friction was highest (small wheels or five layers of coarse cloth) all 100 tests synchronized in antiphase.  For the tests with medium wheel size, 42 of the 100 tests synchronized in phase and 58 synchronized in antiphase. Evidently the basins of attraction for the two states were comparably large under these conditions. Similarly, for the tests with only one layer of coarse cloth, 52 of the 100 tests synchronized in phase, while 48 synchronized in antiphase.

Pantaleone~\cite{pantaleone2002synchronization} also  investigated the effect of increased damping from the platform on the long-term behavior of the system. For the normal set-up (with the metronomes set to different but close frequencies), the metronomes always ended up moving in phase.  However, when the system was placed on a wet surface (increasing the damping on the platform's motion), it was possible to achieve antiphase synchronization.   

\subsection{Stability conditions} 

Our model's predictions are consistent, at least qualitatively, with the experimental findings above. As we will see, when the cubic coefficient $r$ is sufficiently large, the model's behavior  matches the scenario above. At low  damping, the system synchronizes in phase. At high damping, it synchronizes in antiphase. And in between, both types of synchronization are possible. To establish these results, we have derived the conditions for stability of the in-phase and antiphase synchronized states for the general case where platform damping $\mu \neq 0$ and platform restoring stiffness $\kappa \neq 0$. Those conditions are summarized in this subsection.

In the following subsections we will first consider the case where we have nonzero damping of the platform, but allow both $r$ and $\kappa$ to both be zero for simplicity.  We will then explore what happens if $r$ is nonzero, and then $\kappa$.  For nonzero $\mu$ or $\kappa$, the full perturbation analysis is given in Appendix~\ref{s5}.

In these more general cases, the in-phase and antiphase oscillations are found to have amplitude
\begin{eqnarray}
\label{fpa}
\begin{aligned}
A_i =& \sqrt{2} \frac{\theta_c}{\alpha_i} \sqrt{1 +  \sqrt{1 - \alpha_i^2}}, \\
A_a =& \sqrt{2} \frac{\theta_c}{\alpha_a} \sqrt{1 +  \sqrt{1 - \alpha_a^2}},
\end{aligned}
\end{eqnarray}
respectively, where
\begin{eqnarray}
\label{alpha}
\begin{aligned}
\alpha_i =& \left(\frac{\pi\theta_c}{J}\right) \left(\nu + \frac{2 b \mu}{(\kappa-1)^2+\mu^2}\right), \\ \alpha_a =& \frac{\pi\theta_c\nu}{J}.  
\end{aligned} 
\end{eqnarray}
It is interesting that, as before, the steady-state amplitudes $A_i$ and $A_a$ depend on a dimensionless parameter $\alpha$, except now there are two $\alpha$'s, one for in-phase synchronization and another for antiphase. Both $\alpha$'s reflect a balance between damping and driving; the $\nu$ and $\mu$ appearing in the equations above are scaled versions of the pendulum and platform damping, respectively, while $J$ is a scaled impulsive drive strength. Incidentally, note also that the  expression for the antiphase parameter $\alpha_a$ is simpler than that for $\alpha_i$. This makes sense because when antiphase sync occurs, the platform does not move. So that is why the platform parameters $\mu$ and $\kappa$ do not appear in $\alpha_a.$ 

For this general case with $\mu \neq 0$ and $\kappa \neq 0$, the counterparts of the stability boundaries $b_1$ and $b_2$ from Eqs.~\eqref{r-eqs} now become more complicated, but they are still explicitly solvable. Our analysis in Appendix~\ref{stab} shows that the boundaries can be expressed in terms of the following three quantities:
\begin{eqnarray}
\label{sin1}
U = \frac{\nu \pi \theta_c}{J} - \frac{\alpha_i}{1 + \sqrt{1 - \alpha_i^2}}, 
\end{eqnarray}
\begin{eqnarray}
\label{sin2}
V = b - \frac{J \alpha_i}{\pi \theta_c} \frac{\left( \mu \sqrt{1-\alpha_i^2} + (1-\kappa) \alpha_i \right)}{\left( 1 + \sqrt{1-\alpha_i^2} \right)}  \\ \nonumber
+\frac{r \theta_c^2}{4 \alpha_i^2} (1-\kappa) \left( 1 + \sqrt{1-\alpha_i^2} \right),
\end{eqnarray}
and
\begin{eqnarray}
\label{san1}
W = b + \frac{J \alpha_a}{\pi \theta_c} \frac{\left( \mu \sqrt{1-\alpha_a^2} + (1-\kappa) \alpha_a \right)}{\left( 1 + \sqrt{1-\alpha_a^2} \right)}  \\ \nonumber
-\frac{r \theta_c^2}{4 \alpha_a^2} (1-\kappa) \left( 1 + \sqrt{1-\alpha_a^2} \right). 
\end{eqnarray}
The Table~\ref{tab1} summarizes our findings regarding the stability of the in-phase and antiphase fixed points. 

\begin{center}
\begin{table}
\caption{\label{tab1} Stability for the in-phase and antiphase fixed points.}
 \begin{tabular}{|p{2cm} || p{2.75cm} | p{2.75cm} |} 
 \hline
 \textbf{Fixed point} & \textbf{Stable} & \textbf{Unstable}  \\ 
 \hline 
 In-phase &  $U > 0$ and $V>0$ &  $U < 0$ or $V<0$ \\ 
 \hline
 Antiphase & $W>0$ & $W<0$ \\
 \hline
 \end{tabular}
\end{table}
\end{center}

\subsection{Varying the platform damping}

Now, by using the stability criteria above, we can predict what should happen to both forms of synchronization if we vary the platform damping parameter $\mu$. It turns out the results depend qualitatively on the size of the cubic coefficient $r$. As we will see, to match the experimental results of Wu et al.~\cite{wu2012anti} we need to have $r$ sufficiently large, a regime that is plausible for real metronomes. 

\begin{figure}
\centering
\includegraphics[width=2.5in]{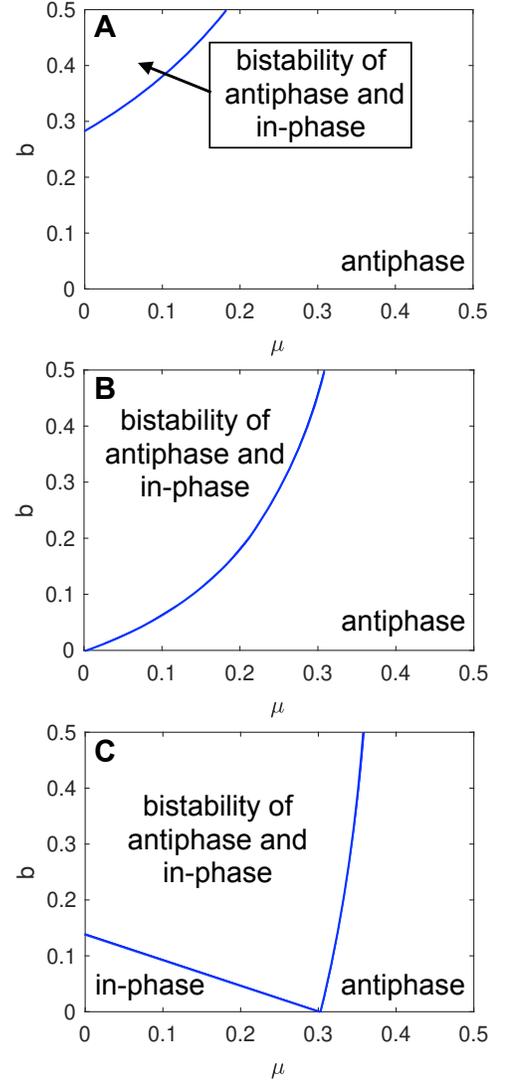} 
\caption{\label{mu_vs_b_k_0} Bifurcation diagram in the $(\mu,b)$ parameter space for three different values of $r$. Recall that $\mu$ measures the dimensionless strength of the damping of the platform, and $b$ measures the back-coupling of the pendulums on the platform; it is proportional to the inertial force (from the swinging of the pendulums) that drives the platform's motion. (A) $r=0$. (B)  $r=r_{c}\approx0.6697$. (C) $r=1 > r_c$. Other parameters: $\theta_c = 0.5$, $J = 3$, $\nu = 1$, $\kappa = 0$. Only the bottom panel (C) is consistent with experimental results on metronomes reported in Wu et al.~\cite{wu2012anti}} 
\end{figure}

Figure~\ref{mu_vs_b_k_0} summarizes the main message of this section. The axes of the parameter space are $\mu$, the dimensionless damping on the platform, and $b$, the dimensionless strength of the inertial driving on the platform due to the swinging of the pendulums. The three panels show what happens as we progressively increase the cubic coefficient $r$. 

In the top two panels, where $r=0$ or $r=r_c$, respectively, there are only two stability regions. In one of them, only the antiphase state is stable. In the other, the antiphase state coexists with a locally stable in-phase state. The important point is that in-phase synchrony is \emph{never} the only attractor here. 

This finding is incompatible with the experimental results of Wu et al.~\cite{wu2012anti} discussed above. They observed a third region, in which the in-phase state became the only attractor. Indeed, they found that their metronomes synchronized in phase \emph{every time} out of 100 tests when the damping was sufficiently low. 

It is only when we get to the bottom panel of Fig.~\ref{mu_vs_b_k_0}, for $r > r_c$, that we could see something like this. In that panel alone, a stability region opens up for low damping $\mu$ in which in-phase synchrony is the only stable state.

\subsubsection{Antiphase sync is stable when $0\leq r\leq r_{c}$ and $\kappa = 0$}

In more detail, the blue curve in the top two panels of Fig.~\ref{mu_vs_b_k_0} is the curve given implicitly by $V=0$. Crossing this curve horizontally by increasing $\mu$ while leaving $b$ fixed corresponds to a pitchfork bifurcation, in which a locally stable in-phase state turns into a saddle equilibrium and branches off two saddle equilibria (all of which are unstable and therefore unobservable in experiments). Not pictured in these figures is the curve given implicitly by $U=0$.  This curve is located further into the first quadrant of the $\mu-b$ plane (larger $\mu$ and larger $b$) and corresponds to a subcritical Hopf bifurcation as the curve is crossed from left to right.

Mathematically, antiphase synchronization is always stable because $W\geq0$, with equality only when $r=r_{c}$. This is easy to see if we rewrite the $ W$ equation (see Eq.~(\ref{san1})) in terms of $r_{c}$, giving
\begin{align}
\label{W-r-cr}
W = & b + \frac{J \alpha_a}{\pi \theta_c} \frac{\left( \mu \sqrt{1-\alpha_a^2} \right)}{\left( 1 + \sqrt{1-\alpha_a^2} \right)}\\ \nonumber
 &+ \frac{\theta_{c}^2(1-\kappa)}{4\alpha_{a}^2}\left(1+\sqrt{1-\alpha_{a}^2}\right)(r_{c}-r).
\end{align}
Observe that when $r<r_{c}$ and $\kappa <1$, each term is positive, so $W$ is also positive. When $r=r_{c}$, $W$ is positive except when $b=\mu=0$.  Because $W\geq0$, we can never see a transition from bistability to only in-phase synchrony.

\subsubsection{Three stability regions when $r> r_{c}$}

In contrast, when the cubic coefficient $r$ is larger than $r_{c}$ (but not too large), there is a regime where only in-phase synchrony is stable. This can be seen in the bottom panel of of Fig.~\ref{mu_vs_b_k_0}, in the lower left corner. This regime occurs when $\mu$ (the damping of the platform) and $b$ (the inertial driving of the platform caused by the swinging of the pendulums) are both sufficiently weak.  

Increasing the damping $\mu$ first stabilizes the antiphase state, when the first bifurcation curve is crossed. Then, when the second curve is crossed, the in-phase oscillations are destabilized and only antiphase synchrony remains stable. 

Full disclosure: For very large $r$, more complicated behavior is possible (but is not shown here, because it is not the object of our interest).

The bifurcation curves shown in Fig.~\ref{mu_vs_b_k_0} can be obtained analytically. When $r> r_{c}$ (as defined by  Eq.~\eqref{r_crit}), the boundary of the global stability region for in-phase synchrony is given by $W=0$. Solving $W=0$ for $b$ in terms of $\mu$ in Eq.~\eqref{W-r-cr} yields an equation for a straight line in the $\mu-b$ plane given by
\begin{align}
\label{b3}
b_3(\mu)=-&\mu\frac{J \alpha_a}{\pi \theta_c} \frac{\left( \sqrt{1-\alpha_a^2} \right)}{\left( 1 + \sqrt{1-\alpha_a^2} \right)}\\ \nonumber
&-\frac{\theta_{c}^2}{4\alpha_{a}^2}(1-\kappa)\left(1+\sqrt{1-\alpha_{a}^2}\right)(r_{c}-r).
 \end{align}
This straight line is depicted in the bottom panel in Fig.~\ref{mu_vs_b_k_0}.  The other bifurcation curve in that panel is defined implicitly by $V=0$.

\subsubsection{Insignificant effect of including $0<\kappa \ll 1$}

\begin{figure}
\centering
\includegraphics[width=2.5in]{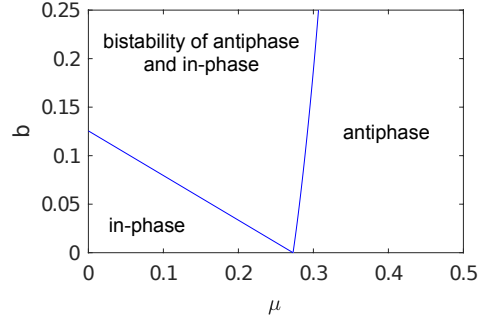} 
\caption{\label{mu_vs_b_nonzero} Bifurcation diagram in the parameters $\mu$ and $b$, when the values of the other parameters are $\theta_c = 0.5$, $J = 3$, $\nu = 1$, $\kappa = 0.1$ and $r=1$. The straight line intersecting the $b$-axis is $b_3(\mu)$ in Eq.~\eqref{b3}.  The other curve is given implicitly by $V=0$.  Both curves correspond to curves of pitchfork bifurcations.} 
\end{figure}

Figure~\ref{mu_vs_b_nonzero} shows that including a small restoring force ($0<\kappa \ll 1$) on the platform does not qualitatively alter the transition scenario described above. For $\kappa<1$ and $r> r_c$, the line $b_3(\mu)$ given by Eq.~\eqref{b3} still passes through the first quadrant of the $\mu-b$ parameter plane, thereby creating a region in which in-phase synchrony is globally stable.  As $\kappa$ approaches $1$, the size of this region shrinks to zero. 

\section{Three-Timescale Analysis}
\label{s8}

We can gain further analytical insight by considering a weak-coupling regime in which $b \ll 1$. Then the indirect coupling from one pendulum onto the other (mediated by the motion of the platform) becomes so weak that the pendulums adjust their phase difference $\psi = \varphi_1 - \varphi_2$ on a  super-long timescale of $t = O(\epsilon^{-1} b^{-1})$. Thanks to this extra separation of timescales, it becomes possible to simplify the slow flow system even more, because in this regime the amplitudes relax to their equilibrium values much more rapidly than the phase difference adjusts. Hence the amplitudes can be adiabatically eliminated, which allows us to reduce the system to a single equation for the phase difference $\psi$, as we will now show. We will work with the general case where $\mu\not=0$ and $\kappa\not=0$, for which the dynamics are described by the full slow flow system~(\ref{as1full}),~(\ref{as3full}) and~(\ref{psifull}), as given in Appendix~\ref{s5}.   

If we were to set $b=0$ in the amplitude equations~(\ref{as1full}) and~(\ref{as3full}), we would find that both $A_1$ and $A_2$ would approach the steady-state amplitude $A_a$ (defined in Eq.~(\ref{fpa})) as the slow time $\tau$ increases. This suggests the ansatz 
\begin{eqnarray}
\nonumber
A_1(\tau) \sim A_a + b a_1(\tau, s ) + \cdots \\ \label{eb}
A_2(\tau) \sim A_a + b a_2(\tau, s ) + \cdots
\end{eqnarray}
in the parameter regime $b\ll 1$, where $a_1$ and $a_2$ are functions of the slow time $\tau$ and a super-slow time $s$, where 
$$s = b \tau.$$ 
Thus, let $a_1 = a_1(\tau,s)$ and $a_2 = a_2(\tau,s)$. If we were to plug the ansatz~(\ref{eb}) into the equation~(\ref{psifull}) for the phase difference $\psi$, we would find that the right hand side of that equation is of $O(b)$. This observation suggests the following ansatz for $\psi$:
\begin{eqnarray}
\label{ebs}
\psi(\tau) \sim \psi_0(s) + b \psi_1(\tau, s) + \cdots,
\end{eqnarray}
where $\psi_0$ is a function of only one variable, $\psi_0 = \psi_0(s)$ but $\psi_1$ is a function of two variables, $\psi_1 = \psi_1(\tau,s)$.

We again carry out a two-timescale analysis. In Eqs.~(\ref{as1full}), (\ref{as3full}), and (\ref{psifull}) we replace $d/d\tau$ by
\begin{eqnarray}
 \frac{\partial}{\partial \tau} + b \frac{\partial }{\partial s},
\end{eqnarray}
plug in the ansatz~(\ref{eb}) and~(\ref{ebs}), and expand in powers of $b$. We find that at first order in $b$,   Eqs.~(\ref{as1full}) and~(\ref{as3full}) reduce to
\begin{eqnarray}
\label{as1b}
\begin{aligned}
\frac{\partial a_1 }{\partial \tau} =& \left( -\frac{\nu}{2} + \frac{J \theta_c^2}{\pi A_a^3} \left(1-\frac{\theta_c^2}{A_a^2} \right)^{-1/2} \right) a_1 + \\ 
&  \frac{\left( (1-\kappa) \sin \psi_0 - (1+\mu) \cos\psi_0 \right)}{2\left((\kappa-1)^2+\mu^2\right)} A_a
\end{aligned}
\end{eqnarray}
and
\begin{eqnarray}
\label{as1c}
\begin{aligned}
\frac{\partial a_2 }{\partial \tau} =& \left( -\frac{\nu}{2} + \frac{J \theta_c^2}{\pi A_a^3} \left(1-\frac{\theta_c^2}{A_a^2} \right)^{-1/2} \right) a_2 + \\ 
&  \frac{\left( (\kappa-1) \sin \psi_0 - (1+\mu) \cos\psi_0 \right)}{2\left((\kappa-1)^2+\mu^2\right)} A_a.
\end{aligned}
\end{eqnarray}

Simple but tedious algebra (that we do not present here) reveals that the quantity in large parentheses above is negative. This result implies that both $a_1$ and $a_2$ relax to constants on the timescale of $\tau$. Since we are interested in much longer timescales (on the super-slow timescale of $\tau/b$) we arrive at the following result for the correction terms to the amplitudes: 
\begin{eqnarray}
\begin{aligned}
a_1 \sim& \left( -\frac{\nu}{2} + \frac{J \theta_c^2}{\pi A_a^3} \left(1-\frac{\theta_c^2}{A_a^2} \right)^{-1/2} \right)^{-1} \times \\ & \frac{\left( (\kappa - 1 ) \sin \psi_0 + (1+\mu) \cos\psi_0 \right)}{2\left((\kappa-1)^2+\mu^2\right)} A_a
\end{aligned}
\end{eqnarray}
and
\begin{eqnarray}
\begin{aligned}
a_2 \sim& \left( -\frac{\nu}{2} + \frac{J \theta_c^2}{\pi A_a^3} \left(1-\frac{\theta_c^2}{A_a^2} \right)^{-1/2} \right)^{-1} \times \\ & \frac{\left( (1 - \kappa) \sin \psi_0 + (1+\mu) \cos\psi_0 \right)}{2\left((\kappa-1)^2+\mu^2\right)} A_a.
\end{aligned}
\end{eqnarray}
Next we plug the ansatz~(\ref{eb}) and~(\ref{ebs}) with $a_1$ and $a_2$ given by the above formula into Eq.~(\ref{psifull}). We obtain, to first order in $b$, that 
\begin{eqnarray}
\begin{aligned}
\frac{d \psi_0}{ds} + \frac{\partial \psi_1}{\partial \tau} =& \frac{(\kappa-1)}{((\kappa-1)^2+\mu^2)} \left( \frac{2 \theta_c J}{\pi A_a^2} - \frac{r A_a^2}{8} \right) \times \\ & \left( -\frac{\nu}{2} + \frac{J \theta_c^2}{\pi A_a^3} \left(1-\frac{\theta_c^2}{A_a^2} \right)^{-1/2} \right)^{-1} \sin\psi_0.
\end{aligned}
\end{eqnarray}
Using the above equation and requiring $\psi_1$ to be bounded in $\tau$ leads to the conclusion that $\psi_1$ is independent of $\tau.$ After substitution of $A_a$ and additional algebra, we finally obtain the desired evolution equation for $\psi_0$:  
\begin{equation}
\label{PsiODE}
\frac{d \psi_0}{ds}= \frac{(1-\kappa)(r_c-r) \gamma}{(\kappa - 1)^2 + \mu^2} \sin \psi_0.\\
\end{equation}
Here $r_c$ is given by Eq.~\eqref{r_crit}, and the constant prefactor $\gamma$ is given by
\begin{equation}
\nonumber
\gamma = \frac{\left(1+\sqrt{1-\alpha^2}\right)^2\sqrt{2-\alpha^2+2\sqrt{1-\alpha^2}}}{2\nu\alpha^2\left[\left(1+\sqrt{1-\alpha^2}\right)\sqrt{2-\alpha^2+2\sqrt{1-\alpha^2}}-\alpha^2\right]},
\end{equation}
where, as before, $\alpha = \alpha_a = \nu\pi\theta_c/J$.  The constant $\gamma$ is well defined and positive for $\alpha$ on the interval $[0,1)$.  

Let us pause to enjoy Eq.~(\ref{PsiODE}). It gives us, in the limit $b \ll 1$, a delightfully simple stability criterion for both the in-phase and antiphase fixed points. When the constant term in front of $\sin\psi_0$ is positive, antiphase oscillations are stable and in-phase oscillations are unstable; when the term is negative, in-phase oscillations are stable and antiphase oscillations are unstable. Equation~(\ref{PsiODE}) also reveals that the damping of the platform cannot change the stability of the fixed points in this regime (because $\mu$ appears only in the denominator which will always be positive). Instead we see that the stability is governed solely by the quantities $r-r_c$ and $1-\kappa$, as follows.  When $\kappa<1$, stable antiphase synchronization and unstable in-phase synchronization occur if $r<r_c$. Conversely, if $r>r_c$, we have stable in-phase synchronization and unstable antiphase synchronization.  

This dichotomy agrees with our earlier analysis in Section~\ref{asymp}. As we saw there, the stability of the in-phase and antiphase fixed points is determined by whether $r<r_c$ or $r>r_c$, for the case where $b$ is very small and hence lies close to the $r$ axis in Fig.~\ref{r_vs_b}.  However, recall that in that case we also assumed $\kappa=0$. Now we see that the dichotomy holds a bit more generally. 



\section{Discussion}
\label{s10}

We have modeled the behavior of two coupled pendulums with deadbeat escapement mechanisms driving their motion.  In our analysis of this system, we focused on a parameter regime that is both physically realistic and analytically tractable: a weak-coupling regime in which the ratio of a pendulum's mass to the mass of the entire system is assumed to be small. In this regime, phase adjustments of the pendulums due to inertial forcing from the platform occur over long times relative to the period of the pendulums. By scaling other physical parameters appropriately, we were able to use a multiple timescale analysis to study ``the sympathy of clocks" in a way that appears simpler than most in the existing literature. It allows us to delineate regions in the parameter space where only in-phase synchronization is stable, or where only antiphase synchronization is stable, or where both are stable. For an example of such a scenario, see the bottom panel of Fig.~\ref{mu_vs_b_k_0} and the surrounding discussion.

One of the unusual features of our approach is that we model the escapement mechanism by using discrete impulses in the form of $\delta$-functions.  Other approaches have used different discontinuous functions to model the escapement~\cite{bennett2002huygens,
dilao2009antiphase,
ramirez2016sympathy,
czolczynski2011two,
fradkov2007synchronization} or continuous functions such as a van der Pol term~\cite{jovanovic2012synchronization,
pantaleone2002synchronization,
willms2017huygens,
ramirez2013synchronization} or some other continuous function which gives self-excitation in the system~\cite{wu2012anti}. Importantly, our impulses provide a boost to the pendulum before it reaches the apex of its swing rather than to push it back in the opposite direction. We believe that this model captures an important aspect of how deadbeat escapements actually work.    

Further, when making the small-angle approximation, we expand $\sin\theta$ past the linear term to include the cubic term.  It is much more common to either take only the first order approximation to sine~\cite{bennett2002huygens,
dilao2009antiphase,
fradkov2007synchronization,
jovanovic2012synchronization,
kumon2002controlled,
pantaleone2002synchronization, 
kuznetsov2007synchronization} so that the analysis is more straightforward, or to avoid a small-angle approximation altogether~\cite{willms2017huygens,wu2012anti,czolczynski2011two} although this choice can cause the analysis to become unwieldy.  We find that including the cubic term is crucial to the dynamics of our model. To wit, Fig.~\ref{mu_vs_b_k_0} demonstrates that if the coefficient $r$ of the cubic term is smaller than a critical value, $r_c$, there is no region of the (platform damping, platform coupling) parameter space where in-phase synchrony is globally stable.

The asymptotic analyses that we have presented are similar in some respects to others in the literature~\cite{bennett2002huygens,
fradkov2007synchronization,
jovanovic2012synchronization,
pantaleone2002synchronization}.  However, while previous analyses have predominantly used the mass ratio $m/M$ as the small parameter, we consider a diverse set of small parameters of the same order of magnitude as this one. As a result, we can clearly tease out the separate roles of the platform damping, the back coupling of the pendulums' motion on the platform, the size of the cubic coefficient, the size of the critical angle, and the size of the impulses from the escapement. We have shown analytically how some of these parameters determine  which mode of synchronization is favored: only in-phase, only antiphase, or the bistability of both. 

Regions of bistability have been found in earlier analytical studies~\cite{kumon2002controlled,czolczynski2011two,jovanovic2012synchronization,dilao2009antiphase,ramirez2016sympathy}. Bistability can also occur in reality; although we are perhaps more accustomed to antiphase synchronization of clocks and in-phase synchronization of metronomes, experimental studies have demonstrated that both kinds of devices can display bistability in certain circumstances~\cite{czolczynski2011two,ramirez2013synchronization,wu2012anti}. 


One of our main results is that the slight dependence of a pendulum's frequency on its amplitude, controlled by the dimensionless parameter $r$, can play an outsized role in the long-term dynamics of coupled clocks and metronomes. Although well known for individual pendulums, this effect has not been emphasized in previous analyses of these coupled systems. Indeed, we suspect that the dynamics of Huygens's clocks have resisted a complete analysis for more than 350 years, precisely because small effects like this can play such a pivotal role. 


\section*{Acknowledgements}

Research of A.N.N. was supported by an NSF Mathematical Sciences Postdoctoral Research Fellowship, Award Number DMS-190288.

\section*{Data Availability}

The data that support the findings of this study are almost all available within the article. Any data that are not available can be found from the corresponding author upon  reasonable request.

\appendix

\section{Asymptotic Analysis of the Scaled  System with $\mu\not=0$ and $\kappa\not=0$}
\label{s5}

From the terms that contain the power $\epsilon^0$ in the expansion of (\ref{equ}), we obtain 
\begin{equation*}
\begin{aligned}
\frac{\partial^2\theta_{10}}{\partial t^2}(t,\tau) &+ \theta_{10}(t,\tau)=0\\
\frac{\partial^2\theta_{20}}{\partial t^2} (t,\tau)&+ \theta_{20}(t,\tau)=0\\
\frac{\partial^2 x_0}{\partial t^2}(t,\tau) &+ \mu \frac{\partial x_0}{\partial t}(t,\tau) + \kappa x_0(t,\tau)  \\ &= -b \left( \frac{\partial^2 \theta_{10}}{\partial t^2}(t,\tau) + \frac{\partial^2\theta_{20}}{\partial t^2}(t,\tau) \right).
\end{aligned}
\end{equation*}
The general solution of these equations is again
\begin{eqnarray}
\nonumber
\theta_{10}(t,\tau) &= A_1(\tau) \, \sin(t+\varphi_1(\tau)) \\
\nonumber
\theta_{20}(t,\tau) &= A_2(\tau) \, \sin(t+\varphi_2(\tau))
\end{eqnarray}
 for the angles of the pendula and now
\begin{eqnarray}
\nonumber
x_0=\frac{b\left( (\kappa-1) (\theta_{10} + \theta_{20}) - \mu \left( \frac{\partial \theta_{10}}{\partial t} +  \frac{\partial \theta_{20}}{\partial t} \right) \right)}{(\kappa-1)^2+\mu^2}
\end{eqnarray}
for the position of the platform. Note that we are implicitly assuming that $(\kappa-1)^2+\mu^2\neq 0$. 

As before, differential equations for the evolution of the slow variables $A_1,A_2,\varphi_1, \varphi_2$ will be obtained at the next order of $\epsilon$ after dealing with the delta-function forcing terms due to the repeated impulses provided by the escapement mechanism.
Note that the $O(\epsilon^1$) terms in the expansion of the system~(\ref{equ}) have the same form as was shown in Section~\ref{asymp} since $\mu$ and $\kappa$ appear only in the $x$ equation which contains no epsilons; however, $\mu$ and $\kappa$ do enter into the equations through the second derivative of $x_0$ which appears in both equations.  To derive the slow flow equations for $A_1,A_2,\psi$, we compute the four integrals as in Section~\ref{asymp} and follow the same subsequent step (dividing the phase equations by the corresponding amplitudes and subtracting the result) to obtain the new equations:
\begin{align}
\label{as1full}
\frac{d A_1 }{d \tau} =& -\frac{\nu}{2} A_1 + \sqrt{1-\frac{\theta_c^2}{A_1^2}} \, \frac{J}{\pi}   - \frac{b \mu}{2 \left( (\kappa-1)^2 + \mu^2 \right)} A_1 \\ 
&- \frac{b \left( (\kappa-1) \sin\psi + \mu \cos\psi \right)}{2 \left( (\kappa-1)^2 + \mu^2 \right)}  A_2 
\nonumber
\\
\label{as3full}
\frac{d A_2 }{d \tau} =& -\frac{\nu}{2} A_2 + \sqrt{1-\frac{\theta_c^2}{A_2^2}} \, \frac{J}{\pi}   - \frac{b \mu}{2 \left( (\kappa-1)^2 + \mu^2 \right)} A_2 \\ 
&+ \frac{b \left( (\kappa-1) \sin\psi - \mu \cos\psi \right)}{2 \left( (\kappa-1)^2 + \mu^2 \right)}  A_1 
\nonumber
\\
\label{psifull}
\frac{d\psi}{d\tau} =& \theta_c \, \frac{J}{\pi} \left(A_2^{-2}-A_1^{-2}\right) + \frac{r}{16}  \left(A_2^{2}-A_1^{2}\right)  \\ 
&+ \frac{b (\kappa - 1)}{2 \left( (\kappa-1)^2 + \mu^2 \right)}
 \left(\frac{A_1}{A_2} - \frac{A_2}{A_1}\right) \cos \psi \nonumber \\ 
 &+\frac{b \mu}{2 \left( (\kappa-1)^2 + \mu^2 \right)}
 \left(\frac{A_1}{A_2} + \frac{A_2}{A_1}\right) \sin \psi. 
\nonumber 
\end{align}

Our asymptotic analysis is again valid for $A_1(\tau) > \theta_c$ and $A_2(\tau) > \theta_c$, meaning that the pendulums' swings are large enough to engage the escapement mechanism at all times.  We refer to the system (\ref{as1full}),  (\ref{as3full}), and (\ref{psifull}) as the \emph{full slow flow}.

\section{Stability analysis}
\label{stab}

Here we calculate the stability of the in-phase synchronized state $\psi=0$, $A_1=A_2=A_i$
and the antiphase synchronized state  $\psi=\pi$, $A_1=A_2=A_a$ by regarding them as fixed points of the system~(\ref{as1full}),~(\ref{as3full}), and (\ref{psifull}). (Recall that $A_i$ and $A_a$ were defined in Eq.~(\ref{fpa})). 

We follow the standard steps. We introduce the functions $F_1(A_1,A_2,\psi)$, $F_2(A_1,A_2,\psi)$ and $F_3(A_1,A_2,\psi)$ so that the system (\ref{as1full}),~(\ref{as3full}) and~(\ref{psifull}) reads
\begin{eqnarray}
\nonumber
\frac{d A_1}{d\tau} &= F_1(A_1,A_2,\psi) \\
\nonumber
\frac{d A_2}{d\tau} &= F_2(A_1,A_2,\psi) \\
\nonumber
\frac{d \psi}{d\tau} &= F_3(A_1,A_2,\psi).
\end{eqnarray}

We compute the matrix of partial derivatives 
\begin{eqnarray}
\nonumber
H = H(A_1,A_2,\psi) = 
\begin{bmatrix}
\frac{\partial F_1}{\partial A_1} &
\frac{\partial F_1}{\partial A_2} &
\frac{\partial F_1}{\partial \psi} \\
\frac{\partial F_2}{\partial A_1} &
\frac{\partial F_2}{\partial A_2} &
\frac{\partial F_2}{\partial \psi} \\
\frac{\partial F_3}{\partial A_1} &
\frac{\partial F_3}{\partial A_2} &
\frac{\partial F_3}{\partial \psi}
\end{bmatrix}.
\end{eqnarray}

We then evaluate the matrix $H$ at each of the two fixed points. We define an index variable $\sigma$ such that $\sigma=1$ if we are considering the in-phase fixed point $\psi = 0$, and $\sigma=-1$ if we are considering the antiphase fixed point $\psi = \pi$. 
Then we can obtain a single set of formulas that apply to both fixed points. For the in-phase fixed point, $\alpha = \alpha_i$ and $\sigma = 1$. For the antiphase fixed point, $\alpha = \alpha_a$ and $\sigma = - 1$. We find that 
\begin{eqnarray}
\nonumber
H  = \left[
\begin{array}{ccc}
h_1 & h_2 & h_3 \\ h_2 & h_1 & -h_3 \\ h_4 & -h_4 & h_5 
\end{array}\right],
\end{eqnarray}
where 
\begin{eqnarray}
\nonumber
\begin{aligned}
h_1 =& \frac{J \alpha^3}{2 \pi \theta_c (1 + \sqrt{1 - \alpha^2})^2} - \frac{1}{2} \left( \nu + \frac{b \mu }{(\kappa-1)^2 + \mu^2} \right) \\
\nonumber
h_2 =& - \frac{\sigma  b \mu}{2\left((\kappa-1)^2 + \mu^2\right)} 
\\
\nonumber
h_3 =& - \frac{\sigma  b (\kappa - 1) \theta_c}{\sqrt{2} \alpha \left((\kappa-1)^2 + \mu^2\right)} \sqrt{1 + \sqrt{1 - \alpha^2}}
\\
\nonumber
h_4 =& \frac{J \alpha^3}{\pi \sqrt{2} \theta_c^2} \left( 1 + \sqrt{1 - \alpha^2} \right)^{-3/2} - \frac{r\theta_c}{4\sqrt{2}\alpha} \left( 1 + \sqrt{1 - \alpha^2} \right)^{1/2} \\
&+ \frac{\sigma  b (\kappa - 1) \alpha}{\sqrt{2} \theta_ c \left((\kappa-1)^2 + \mu^2\right)} \left( 1 + \sqrt{1 - \alpha^2} \right)^{-1/2}
\\
\nonumber
h_5 =& \frac{\sigma  b \mu}{(\kappa-1)^2 + \mu^2}. \end{aligned}
\end{eqnarray}

The characteristic polynomial of $H$ is 
\begin{eqnarray}
\begin{aligned}
P(\lambda) =& \left(h_1+h_2 - \lambda \right) \\ &
\left( \lambda^2 - (h_5+h_1-h_2) \lambda + h_5 h_1 - h_5 h_2 - 2 h_3 h_4 \right).
\end{aligned}
\nonumber
\end{eqnarray}
One root of $P(\lambda)$ is $h_1+h_2$. Simple algebra leads to $h_1+h_2 = -J\alpha\sqrt{1-\alpha^2}/(\pi\theta_c(1+\sqrt{1-\alpha^2}))$. Thus, we have that this root is always negative. So the stability of the fixed point is determined by the two other roots. The fixed point will be stable when $h_5+h_1-h_2<0$, and $h_5 h_1 - h_5 h_2 - 2 h_3 h_4>0$. After some algebra we find that this set of inequalities  translates to
\begin{eqnarray}
\nonumber
\frac{2 \sigma b \mu}{(\kappa-1)^2 + \mu^2} -  \frac{J \alpha}{\pi \theta_c} \frac{\sqrt{1-\alpha^2}}{\left( 1+\sqrt{1-\alpha^2} \right)} < 0
\end{eqnarray}
and 
\begin{eqnarray}
\nonumber
b - \frac{\sigma  J \alpha}{\pi \theta_c} \frac{\left( \mu \sqrt{1-\alpha^2} + (1-\kappa) \alpha \right)}{\left( 1 + \sqrt{1-\alpha^2} \right)} + \\ \nonumber
\frac{\sigma r \theta_c^2}{4 \alpha^2} (1-\kappa) \left( 1 + \sqrt{1-\alpha^2} \right) > 0. 
\end{eqnarray}
The first of the above equations is satisfied for all values of the parameters when $\sigma=-1$, i.e. for the antiphase fixed point. This observation and the above formulas lead to the conditions for stability summarized in Section~\ref{s7}.

\bibliography{refe}

\begin{thebibliography}{47}%
\makeatletter
\providecommand \@ifxundefined [1]{%
 \@ifx{#1\undefined}
}%
\providecommand \@ifnum [1]{%
 \ifnum #1\expandafter \@firstoftwo
 \else \expandafter \@secondoftwo
 \fi
}%
\providecommand \@ifx [1]{%
 \ifx #1\expandafter \@firstoftwo
 \else \expandafter \@secondoftwo
 \fi
}%
\providecommand \natexlab [1]{#1}%
\providecommand \enquote  [1]{``#1''}%
\providecommand \bibnamefont  [1]{#1}%
\providecommand \bibfnamefont [1]{#1}%
\providecommand \citenamefont [1]{#1}%
\providecommand \href@noop [0]{\@secondoftwo}%
\providecommand \href [0]{\begingroup \@sanitize@url \@href}%
\providecommand \@href[1]{\@@startlink{#1}\@@href}%
\providecommand \@@href[1]{\endgroup#1\@@endlink}%
\providecommand \@sanitize@url [0]{\catcode `\\12\catcode `\$12\catcode
  `\&12\catcode `\#12\catcode `\^12\catcode `\_12\catcode `\%12\relax}%
\providecommand \@@startlink[1]{}%
\providecommand \@@endlink[0]{}%
\providecommand \url  [0]{\begingroup\@sanitize@url \@url }%
\providecommand \@url [1]{\endgroup\@href {#1}{\urlprefix }}%
\providecommand \urlprefix  [0]{URL }%
\providecommand \Eprint [0]{\href }%
\providecommand \doibase [0]{http://dx.doi.org/}%
\providecommand \selectlanguage [0]{\@gobble}%
\providecommand \bibinfo  [0]{\@secondoftwo}%
\providecommand \bibfield  [0]{\@secondoftwo}%
\providecommand \translation [1]{[#1]}%
\providecommand \BibitemOpen [0]{}%
\providecommand \bibitemStop [0]{}%
\providecommand \bibitemNoStop [0]{.\EOS\space}%
\providecommand \EOS [0]{\spacefactor3000\relax}%
\providecommand \BibitemShut  [1]{\csname bibitem#1\endcsname}%
\let\auto@bib@innerbib\@empty
\bibitem [{\citenamefont {Winfree}(1980)}]{Winfree1980geometry}%
  \BibitemOpen
  \bibfield  {author} {\bibinfo {author} {\bibfnamefont {A.~T.}\ \bibnamefont
  {Winfree}},\ }\href@noop {} {\emph {\bibinfo {title} {The Geometry of
  Biological Time}}}\ (\bibinfo  {publisher} {Springer-Verlag},\ \bibinfo
  {year} {1980})\BibitemShut {NoStop}%
\bibitem [{\citenamefont {Pikovsky}, \citenamefont {Rosenblum},\ and\
  \citenamefont {Kurths}(2001)}]{pikovsky2001synchronization}%
  \BibitemOpen
  \bibfield  {author} {\bibinfo {author} {\bibfnamefont {A.}~\bibnamefont
  {Pikovsky}}, \bibinfo {author} {\bibfnamefont {M.}~\bibnamefont {Rosenblum}},
  \ and\ \bibinfo {author} {\bibfnamefont {J.}~\bibnamefont {Kurths}},\
  }\href@noop {} {\emph {\bibinfo {title} {Synchronization: A Universal Concept
  in Nonlinear Sciences}}}\ (\bibinfo  {publisher} {Cambridge University
  Press},\ \bibinfo {year} {2001})\BibitemShut {NoStop}%
\bibitem [{\citenamefont {Strogatz}(2003)}]{strogatz2003sync}%
  \BibitemOpen
  \bibfield  {author} {\bibinfo {author} {\bibfnamefont {S.~H.}\ \bibnamefont
  {Strogatz}},\ }\href@noop {} {\emph {\bibinfo {title} {Sync}}}\ (\bibinfo
  {publisher} {Hyperion},\ \bibinfo {year} {2003})\BibitemShut {NoStop}%
\bibitem [{\citenamefont {Blekhman}(1988)}]{blekhmansynchronization}%
  \BibitemOpen
  \bibfield  {author} {\bibinfo {author} {\bibfnamefont {I.~I.}\ \bibnamefont
  {Blekhman}},\ }\href@noop {} {\emph {\bibinfo {title} {Synchronization in
  Science and Technology}}}\ (\bibinfo  {publisher} {American Society of
  Mechanical Engineers Press},\ \bibinfo {year} {1988})\BibitemShut {NoStop}%
\bibitem [{\citenamefont {Huygens}(1893)}]{huygens1893oeuvres}%
  \BibitemOpen
  \bibfield  {author} {\bibinfo {author} {\bibfnamefont {C.}~\bibnamefont
  {Huygens}},\ }\href@noop {} {\emph {\bibinfo {title} {{Oeuvres compl{\`e}tes
  de Christiaan Huygens}}}},\ edited by\ \bibinfo {editor} {\bibfnamefont
  {M.}~\bibnamefont {Nijhoff}},\ Vol.~\bibinfo {volume} {5}\ (\bibinfo
  {publisher} {Societ{\'e} Hollandaise des Sciences},\ \bibinfo {year} {1893})\
  pp.\ \bibinfo {pages} {241--262}\BibitemShut {NoStop}%
\bibitem [{\citenamefont {Yoder}(2004)}]{yoder2004unrolling}%
  \BibitemOpen
  \bibfield  {author} {\bibinfo {author} {\bibfnamefont {J.~G.}\ \bibnamefont
  {Yoder}},\ }\href@noop {} {\emph {\bibinfo {title} {{Unrolling Time:
  Christiaan Huygens and the Mathematization of Nature}}}}\ (\bibinfo
  {publisher} {Cambridge University Press},\ \bibinfo {year}
  {2004})\BibitemShut {NoStop}%
\bibitem [{\citenamefont {Ramirez}\ and\ \citenamefont
  {Nijmeijer}(2020)}]{ramirez2020secret}%
  \BibitemOpen
  \bibfield  {author} {\bibinfo {author} {\bibfnamefont {J.~P.}\ \bibnamefont
  {Ramirez}}\ and\ \bibinfo {author} {\bibfnamefont {H.}~\bibnamefont
  {Nijmeijer}},\ }\bibfield  {title} {\enquote {\bibinfo {title} {The secret of
  the synchronized pendulums},}\ }\href@noop {} {\bibfield  {journal} {\bibinfo
   {journal} {Physics World}\ }\textbf {\bibinfo {volume} {33}},\ \bibinfo
  {pages} {36} (\bibinfo {year} {2020})}\BibitemShut {NoStop}%
\bibitem [{\citenamefont {Ellicott}(1740{\natexlab{a}})}]{ellicott1740acount}%
  \BibitemOpen
  \bibfield  {author} {\bibinfo {author} {\bibfnamefont {J.}~\bibnamefont
  {Ellicott}},\ }\bibfield  {title} {\enquote {\bibinfo {title} {An account of
  the influence which two pendulum clocks were observed to have upon each
  other},}\ }\href@noop {} {\bibfield  {journal} {\bibinfo  {journal} {Phil
  Trans R Soc}\ }\textbf {\bibinfo {volume} {41}},\ \bibinfo {pages} {126--128}
  (\bibinfo {year} {1740}{\natexlab{a}})}\BibitemShut {NoStop}%
\bibitem [{\citenamefont {Ellicott}(1740{\natexlab{b}})}]{ellicott1740further}%
  \BibitemOpen
  \bibfield  {author} {\bibinfo {author} {\bibfnamefont {J.}~\bibnamefont
  {Ellicott}},\ }\bibfield  {title} {\enquote {\bibinfo {title} {Further
  observations and experiments concerning the two clocks above mentioned},}\
  }\href@noop {} {\bibfield  {journal} {\bibinfo  {journal} {Phil Trans R Soc}\
  }\textbf {\bibinfo {volume} {41}},\ \bibinfo {pages} {128--135} (\bibinfo
  {year} {1740}{\natexlab{b}})}\BibitemShut {NoStop}%
\bibitem [{\citenamefont {Ellis}(1873)}]{ellis1873sympathetic}%
  \BibitemOpen
  \bibfield  {author} {\bibinfo {author} {\bibfnamefont {W.}~\bibnamefont
  {Ellis}},\ }\bibfield  {title} {\enquote {\bibinfo {title} {On sympathetic
  influence between clocks},}\ }\href@noop {} {\bibfield  {journal} {\bibinfo
  {journal} {Monthly Notices of the Royal Astronomical Society}\ }\textbf
  {\bibinfo {volume} {33}},\ \bibinfo {pages} {480} (\bibinfo {year}
  {1873})}\BibitemShut {NoStop}%
\bibitem [{\citenamefont {Korteweg}(1906)}]{korteweg1906horloges}%
  \BibitemOpen
  \bibfield  {author} {\bibinfo {author} {\bibfnamefont {D.}~\bibnamefont
  {Korteweg}},\ }\bibfield  {title} {\enquote {\bibinfo {title} {{Les horloges
  sympathiques de Huygens}},}\ }\href@noop {} {\bibfield  {journal} {\bibinfo
  {journal} {Archives Neerlandaises, Serie II}\ }\textbf {\bibinfo {volume}
  {11}},\ \bibinfo {pages} {273--295} (\bibinfo {year} {1906})}\BibitemShut
  {NoStop}%
\bibitem [{\citenamefont {Bennett}\ \emph {et~al.}(2002)\citenamefont
  {Bennett}, \citenamefont {Schatz}, \citenamefont {Rockwood},\ and\
  \citenamefont {Wiesenfeld}}]{bennett2002huygens}%
  \BibitemOpen
  \bibfield  {author} {\bibinfo {author} {\bibfnamefont {M.}~\bibnamefont
  {Bennett}}, \bibinfo {author} {\bibfnamefont {M.~F.}\ \bibnamefont {Schatz}},
  \bibinfo {author} {\bibfnamefont {H.}~\bibnamefont {Rockwood}}, \ and\
  \bibinfo {author} {\bibfnamefont {K.}~\bibnamefont {Wiesenfeld}},\ }\bibfield
   {title} {\enquote {\bibinfo {title} {{Huygens's clocks}},}\ }\href@noop {}
  {\bibfield  {journal} {\bibinfo  {journal} {Proc R Soc A}\ }\textbf {\bibinfo
  {volume} {458}},\ \bibinfo {pages} {563--579} (\bibinfo {year}
  {2002})}\BibitemShut {NoStop}%
\bibitem [{\citenamefont {Oud}, \citenamefont {Nijmeijer},\ and\ \citenamefont
  {Pogromsky}(2006)}]{oud2006study}%
  \BibitemOpen
  \bibfield  {author} {\bibinfo {author} {\bibfnamefont {W.}~\bibnamefont
  {Oud}}, \bibinfo {author} {\bibfnamefont {H.}~\bibnamefont {Nijmeijer}}, \
  and\ \bibinfo {author} {\bibfnamefont {A.~Y.}\ \bibnamefont {Pogromsky}},\
  }\bibfield  {title} {\enquote {\bibinfo {title} {{A study of Huijgens’
  synchronization: experimental results}},}\ }in\ \href@noop {} {\emph
  {\bibinfo {booktitle} {Group Coordination and Cooperative Control}}}\
  (\bibinfo  {publisher} {Springer},\ \bibinfo {year} {2006})\ pp.\ \bibinfo
  {pages} {191--203}\BibitemShut {NoStop}%
\bibitem [{\citenamefont {Senator}(2006)}]{senator2006synchronization}%
  \BibitemOpen
  \bibfield  {author} {\bibinfo {author} {\bibfnamefont {M.}~\bibnamefont
  {Senator}},\ }\bibfield  {title} {\enquote {\bibinfo {title} {Synchronization
  of two coupled escapement-driven pendulum clocks},}\ }\href@noop {}
  {\bibfield  {journal} {\bibinfo  {journal} {J Sound Vib}\ }\textbf {\bibinfo
  {volume} {291}},\ \bibinfo {pages} {566--603} (\bibinfo {year}
  {2006})}\BibitemShut {NoStop}%
\bibitem [{\citenamefont {Dil{\~a}o}(2009)}]{dilao2009antiphase}%
  \BibitemOpen
  \bibfield  {author} {\bibinfo {author} {\bibfnamefont {R.}~\bibnamefont
  {Dil{\~a}o}},\ }\bibfield  {title} {\enquote {\bibinfo {title} {{Antiphase
  and in-phase synchronization of nonlinear oscillators: The Huygens's clocks
  system}},}\ }\href@noop {} {\bibfield  {journal} {\bibinfo  {journal}
  {Chaos}\ }\textbf {\bibinfo {volume} {19}},\ \bibinfo {pages} {023118}
  (\bibinfo {year} {2009})}\BibitemShut {NoStop}%
\bibitem [{\citenamefont {Czolczynski}\ \emph {et~al.}(2009)\citenamefont
  {Czolczynski}, \citenamefont {Perlikowski}, \citenamefont {Stefanski},\ and\
  \citenamefont {Kapitaniak}}]{czolczynski2009clustering}%
  \BibitemOpen
  \bibfield  {author} {\bibinfo {author} {\bibfnamefont {K.}~\bibnamefont
  {Czolczynski}}, \bibinfo {author} {\bibfnamefont {P.}~\bibnamefont
  {Perlikowski}}, \bibinfo {author} {\bibfnamefont {A.}~\bibnamefont
  {Stefanski}}, \ and\ \bibinfo {author} {\bibfnamefont {T.}~\bibnamefont
  {Kapitaniak}},\ }\bibfield  {title} {\enquote {\bibinfo {title} {{Clustering
  of Huygens' clocks}},}\ }\href@noop {} {\bibfield  {journal} {\bibinfo
  {journal} {Prog Theo Phys}\ }\textbf {\bibinfo {volume} {122}},\ \bibinfo
  {pages} {1027--1033} (\bibinfo {year} {2009})}\BibitemShut {NoStop}%
\bibitem [{\citenamefont {Czo{\l}czy{\'n}ski}\ \emph
  {et~al.}(2011)\citenamefont {Czo{\l}czy{\'n}ski}, \citenamefont
  {Perlikowski}, \citenamefont {Stefa{\'n}ski},\ and\ \citenamefont
  {Kapitaniak}}]{czolczynski2011two}%
  \BibitemOpen
  \bibfield  {author} {\bibinfo {author} {\bibfnamefont {K.}~\bibnamefont
  {Czo{\l}czy{\'n}ski}}, \bibinfo {author} {\bibfnamefont {P.}~\bibnamefont
  {Perlikowski}}, \bibinfo {author} {\bibfnamefont {A.}~\bibnamefont
  {Stefa{\'n}ski}}, \ and\ \bibinfo {author} {\bibfnamefont {T.}~\bibnamefont
  {Kapitaniak}},\ }\bibfield  {title} {\enquote {\bibinfo {title} {Why two
  clocks synchronize: Energy balance of the synchronized clocks},}\ }\href@noop
  {} {\bibfield  {journal} {\bibinfo  {journal} {Chaos}\ }\textbf {\bibinfo
  {volume} {21}},\ \bibinfo {pages} {023129} (\bibinfo {year}
  {2011})}\BibitemShut {NoStop}%
\bibitem [{\citenamefont {Czolczynski}\ \emph {et~al.}(2011)\citenamefont
  {Czolczynski}, \citenamefont {Perlikowski}, \citenamefont {Stefanski},\ and\
  \citenamefont {Kapitaniak}}]{czolczynski2011huygens}%
  \BibitemOpen
  \bibfield  {author} {\bibinfo {author} {\bibfnamefont {K.}~\bibnamefont
  {Czolczynski}}, \bibinfo {author} {\bibfnamefont {P.}~\bibnamefont
  {Perlikowski}}, \bibinfo {author} {\bibfnamefont {A.}~\bibnamefont
  {Stefanski}}, \ and\ \bibinfo {author} {\bibfnamefont {T.}~\bibnamefont
  {Kapitaniak}},\ }\bibfield  {title} {\enquote {\bibinfo {title} {Huygens' odd
  sympathy experiment revisited},}\ }\href@noop {} {\bibfield  {journal}
  {\bibinfo  {journal} {International Journal of Bifurcation and Chaos}\
  }\textbf {\bibinfo {volume} {21}},\ \bibinfo {pages} {2047--2056} (\bibinfo
  {year} {2011})}\BibitemShut {NoStop}%
\bibitem [{\citenamefont {Czolczynski}\ \emph {et~al.}(2013)\citenamefont
  {Czolczynski}, \citenamefont {Perlikowski}, \citenamefont {Stefanski},\ and\
  \citenamefont {Kapitaniak}}]{czolczynski2013synchronization}%
  \BibitemOpen
  \bibfield  {author} {\bibinfo {author} {\bibfnamefont {K.}~\bibnamefont
  {Czolczynski}}, \bibinfo {author} {\bibfnamefont {P.}~\bibnamefont
  {Perlikowski}}, \bibinfo {author} {\bibfnamefont {A.}~\bibnamefont
  {Stefanski}}, \ and\ \bibinfo {author} {\bibfnamefont {T.}~\bibnamefont
  {Kapitaniak}},\ }\bibfield  {title} {\enquote {\bibinfo {title}
  {Synchronization of the self-excited pendula suspended on the vertically
  displacing beam},}\ }\href@noop {} {\bibfield  {journal} {\bibinfo  {journal}
  {Communications in Nonlinear Science and Numerical Simulation}\ }\textbf
  {\bibinfo {volume} {18}},\ \bibinfo {pages} {386--400} (\bibinfo {year}
  {2013})}\BibitemShut {NoStop}%
\bibitem [{\citenamefont {Kapitaniak}\ \emph {et~al.}(2012)\citenamefont
  {Kapitaniak}, \citenamefont {Czolczynski}, \citenamefont {Perlikowski},
  \citenamefont {Stefanski},\ and\ \citenamefont
  {Kapitaniak}}]{kapitaniak2012synchronization}%
  \BibitemOpen
  \bibfield  {author} {\bibinfo {author} {\bibfnamefont {M.}~\bibnamefont
  {Kapitaniak}}, \bibinfo {author} {\bibfnamefont {K.}~\bibnamefont
  {Czolczynski}}, \bibinfo {author} {\bibfnamefont {P.}~\bibnamefont
  {Perlikowski}}, \bibinfo {author} {\bibfnamefont {A.}~\bibnamefont
  {Stefanski}}, \ and\ \bibinfo {author} {\bibfnamefont {T.}~\bibnamefont
  {Kapitaniak}},\ }\bibfield  {title} {\enquote {\bibinfo {title}
  {Synchronization of clocks},}\ }\href@noop {} {\bibfield  {journal} {\bibinfo
   {journal} {Physics Reports}\ }\textbf {\bibinfo {volume} {517}},\ \bibinfo
  {pages} {1--69} (\bibinfo {year} {2012})}\BibitemShut {NoStop}%
\bibitem [{\citenamefont {Jovanovic}\ and\ \citenamefont
  {Koshkin}(2012)}]{jovanovic2012synchronization}%
  \BibitemOpen
  \bibfield  {author} {\bibinfo {author} {\bibfnamefont {V.}~\bibnamefont
  {Jovanovic}}\ and\ \bibinfo {author} {\bibfnamefont {S.}~\bibnamefont
  {Koshkin}},\ }\bibfield  {title} {\enquote {\bibinfo {title}
  {{Synchronization of Huygens' clocks and the Poincar{\'e} method}},}\
  }\href@noop {} {\bibfield  {journal} {\bibinfo  {journal} {J Sound Vib}\
  }\textbf {\bibinfo {volume} {331}},\ \bibinfo {pages} {2887--2900} (\bibinfo
  {year} {2012})}\BibitemShut {NoStop}%
\bibitem [{\citenamefont {Ramirez}, \citenamefont {Fey},\ and\ \citenamefont
  {Nijmeijer}(2013)}]{ramirez2013synchronization}%
  \BibitemOpen
  \bibfield  {author} {\bibinfo {author} {\bibfnamefont {J.~P.}\ \bibnamefont
  {Ramirez}}, \bibinfo {author} {\bibfnamefont {R.~H.}\ \bibnamefont {Fey}}, \
  and\ \bibinfo {author} {\bibfnamefont {H.}~\bibnamefont {Nijmeijer}},\
  }\bibfield  {title} {\enquote {\bibinfo {title} {{Synchronization of weakly
  nonlinear oscillators with Huygens' coupling}},}\ }\href@noop {} {\bibfield
  {journal} {\bibinfo  {journal} {Chaos}\ }\textbf {\bibinfo {volume} {23}},\
  \bibinfo {pages} {033118} (\bibinfo {year} {2013})}\BibitemShut {NoStop}%
\bibitem [{\citenamefont {Ramirez}\ \emph
  {et~al.}(2014{\natexlab{a}})\citenamefont {Ramirez}, \citenamefont {Fey},
  \citenamefont {Aihara},\ and\ \citenamefont
  {Nijmeijer}}]{ramirez2014improved}%
  \BibitemOpen
  \bibfield  {author} {\bibinfo {author} {\bibfnamefont {J.~P.}\ \bibnamefont
  {Ramirez}}, \bibinfo {author} {\bibfnamefont {R.}~\bibnamefont {Fey}},
  \bibinfo {author} {\bibfnamefont {K.}~\bibnamefont {Aihara}}, \ and\ \bibinfo
  {author} {\bibfnamefont {H.}~\bibnamefont {Nijmeijer}},\ }\bibfield  {title}
  {\enquote {\bibinfo {title} {{An improved model for the classical Huygens'
  experiment on synchronization of pendulum clocks}},}\ }\href@noop {}
  {\bibfield  {journal} {\bibinfo  {journal} {Journal of Sound and Vibration}\
  }\textbf {\bibinfo {volume} {333}},\ \bibinfo {pages} {7248--7266} (\bibinfo
  {year} {2014}{\natexlab{a}})}\BibitemShut {NoStop}%
\bibitem [{\citenamefont {Ramirez}\ \emph
  {et~al.}(2014{\natexlab{b}})\citenamefont {Ramirez}, \citenamefont {Aihara},
  \citenamefont {Fey},\ and\ \citenamefont {Nijmeijer}}]{ramirez2014further}%
  \BibitemOpen
  \bibfield  {author} {\bibinfo {author} {\bibfnamefont {J.~P.}\ \bibnamefont
  {Ramirez}}, \bibinfo {author} {\bibfnamefont {K.}~\bibnamefont {Aihara}},
  \bibinfo {author} {\bibfnamefont {R.}~\bibnamefont {Fey}}, \ and\ \bibinfo
  {author} {\bibfnamefont {H.}~\bibnamefont {Nijmeijer}},\ }\bibfield  {title}
  {\enquote {\bibinfo {title} {{Further understanding of Huygens' coupled
  clocks: The effect of stiffness}},}\ }\href@noop {} {\bibfield  {journal}
  {\bibinfo  {journal} {Physica D: Nonlinear Phenomena}\ }\textbf {\bibinfo
  {volume} {270}},\ \bibinfo {pages} {11--19} (\bibinfo {year}
  {2014}{\natexlab{b}})}\BibitemShut {NoStop}%
\bibitem [{\citenamefont {Ramirez}\ \emph {et~al.}(2016)\citenamefont
  {Ramirez}, \citenamefont {Olvera}, \citenamefont {Nijmeijer},\ and\
  \citenamefont {Alvarez}}]{ramirez2016sympathy}%
  \BibitemOpen
  \bibfield  {author} {\bibinfo {author} {\bibfnamefont {J.~P.}\ \bibnamefont
  {Ramirez}}, \bibinfo {author} {\bibfnamefont {L.~A.}\ \bibnamefont {Olvera}},
  \bibinfo {author} {\bibfnamefont {H.}~\bibnamefont {Nijmeijer}}, \ and\
  \bibinfo {author} {\bibfnamefont {J.}~\bibnamefont {Alvarez}},\ }\bibfield
  {title} {\enquote {\bibinfo {title} {{The sympathy of two pendulum clocks:
  beyond Huygens' observations}},}\ }\href@noop {} {\bibfield  {journal}
  {\bibinfo  {journal} {Scientific Reports}\ }\textbf {\bibinfo {volume} {6}},\
  \bibinfo {pages} {23580} (\bibinfo {year} {2016})}\BibitemShut {NoStop}%
\bibitem [{\citenamefont {Ramirez}\ and\ \citenamefont
  {Nijmeijer}(2016)}]{ramirez2016poincare}%
  \BibitemOpen
  \bibfield  {author} {\bibinfo {author} {\bibfnamefont {J.~P.}\ \bibnamefont
  {Ramirez}}\ and\ \bibinfo {author} {\bibfnamefont {H.}~\bibnamefont
  {Nijmeijer}},\ }\bibfield  {title} {\enquote {\bibinfo {title} {{The
  Poincar{\'e} method: a powerful tool for analyzing synchronization of coupled
  oscillators}},}\ }\href@noop {} {\bibfield  {journal} {\bibinfo  {journal}
  {Indagationes Mathematicae}\ }\textbf {\bibinfo {volume} {27}},\ \bibinfo
  {pages} {1127--1146} (\bibinfo {year} {2016})}\BibitemShut {NoStop}%
\bibitem [{\citenamefont {Oliveira}\ and\ \citenamefont
  {Melo}(2015)}]{oliveira2015huygens}%
  \BibitemOpen
  \bibfield  {author} {\bibinfo {author} {\bibfnamefont {H.~M.}\ \bibnamefont
  {Oliveira}}\ and\ \bibinfo {author} {\bibfnamefont {L.~V.}\ \bibnamefont
  {Melo}},\ }\bibfield  {title} {\enquote {\bibinfo {title} {{Huygens
  synchronization of two clocks}},}\ }\href@noop {} {\bibfield  {journal}
  {\bibinfo  {journal} {Scientific reports}\ }\textbf {\bibinfo {volume} {5}},\
  \bibinfo {pages} {11548} (\bibinfo {year} {2015})}\BibitemShut {NoStop}%
\bibitem [{\citenamefont {Willms}, \citenamefont {Kitanov},\ and\ \citenamefont
  {Langford}(2017)}]{willms2017huygens}%
  \BibitemOpen
  \bibfield  {author} {\bibinfo {author} {\bibfnamefont {A.~R.}\ \bibnamefont
  {Willms}}, \bibinfo {author} {\bibfnamefont {P.~M.}\ \bibnamefont {Kitanov}},
  \ and\ \bibinfo {author} {\bibfnamefont {W.~F.}\ \bibnamefont {Langford}},\
  }\bibfield  {title} {\enquote {\bibinfo {title} {{Huygens' clocks
  revisited}},}\ }\href@noop {} {\bibfield  {journal} {\bibinfo  {journal}
  {Royal Society Open Science}\ }\textbf {\bibinfo {volume} {4}},\ \bibinfo
  {pages} {170777} (\bibinfo {year} {2017})}\BibitemShut {NoStop}%
\bibitem [{\citenamefont {Wiesenfeld}(2017)}]{wiesenfeld2017huygens}%
  \BibitemOpen
  \bibfield  {author} {\bibinfo {author} {\bibfnamefont {K.}~\bibnamefont
  {Wiesenfeld}},\ }\bibfield  {title} {\enquote {\bibinfo {title} {{Huygens's
  odd sympathy recreated}},}\ }\href@noop {} {\bibfield  {journal} {\bibinfo
  {journal} {Societate si Politica}\ }\textbf {\bibinfo {volume} {11}},\
  \bibinfo {pages} {15--22} (\bibinfo {year} {2017})}\BibitemShut {NoStop}%
\bibitem [{\citenamefont {Pantaleone}(2002)}]{pantaleone2002synchronization}%
  \BibitemOpen
  \bibfield  {author} {\bibinfo {author} {\bibfnamefont {J.}~\bibnamefont
  {Pantaleone}},\ }\bibfield  {title} {\enquote {\bibinfo {title}
  {Synchronization of metronomes},}\ }\href@noop {} {\bibfield  {journal}
  {\bibinfo  {journal} {Am J Phys}\ }\textbf {\bibinfo {volume} {70}},\
  \bibinfo {pages} {992--1000} (\bibinfo {year} {2002})}\BibitemShut {NoStop}%
\bibitem [{\citenamefont {Kuznetsov}\ \emph {et~al.}(2007)\citenamefont
  {Kuznetsov}, \citenamefont {Leonov}, \citenamefont {Nijmeijer},\ and\
  \citenamefont {Pogromsky}}]{kuznetsov2007synchronization}%
  \BibitemOpen
  \bibfield  {author} {\bibinfo {author} {\bibfnamefont {N.~V.}\ \bibnamefont
  {Kuznetsov}}, \bibinfo {author} {\bibfnamefont {G.~A.}\ \bibnamefont
  {Leonov}}, \bibinfo {author} {\bibfnamefont {H.}~\bibnamefont {Nijmeijer}}, \
  and\ \bibinfo {author} {\bibfnamefont {A.~Y.}\ \bibnamefont {Pogromsky}},\
  }\bibfield  {title} {\enquote {\bibinfo {title} {Synchronization of two
  metronomes},}\ }in\ \href@noop {} {\emph {\bibinfo {booktitle} {Proceedings
  of the 3rd IFAC Workshop (PSYCO'07), 29-31 August 2007, Saint Petersburg,
  Russia}}}\ (\bibinfo {year} {2007})\ pp.\ \bibinfo {pages}
  {49--52}\BibitemShut {NoStop}%
\bibitem [{\citenamefont {Ulrichs}, \citenamefont {Mann},\ and\ \citenamefont
  {Parlitz}(2009)}]{ulrichs2009synchronization}%
  \BibitemOpen
  \bibfield  {author} {\bibinfo {author} {\bibfnamefont {H.}~\bibnamefont
  {Ulrichs}}, \bibinfo {author} {\bibfnamefont {A.}~\bibnamefont {Mann}}, \
  and\ \bibinfo {author} {\bibfnamefont {U.}~\bibnamefont {Parlitz}},\
  }\bibfield  {title} {\enquote {\bibinfo {title} {Synchronization and chaotic
  dynamics of coupled mechanical metronomes},}\ }\href@noop {} {\bibfield
  {journal} {\bibinfo  {journal} {Chaos}\ }\textbf {\bibinfo {volume} {19}},\
  \bibinfo {pages} {043120} (\bibinfo {year} {2009})}\BibitemShut {NoStop}%
\bibitem [{\citenamefont {Wu}\ \emph {et~al.}(2012)\citenamefont {Wu},
  \citenamefont {Wang}, \citenamefont {Li},\ and\ \citenamefont
  {Xiao}}]{wu2012anti}%
  \BibitemOpen
  \bibfield  {author} {\bibinfo {author} {\bibfnamefont {Y.}~\bibnamefont
  {Wu}}, \bibinfo {author} {\bibfnamefont {N.}~\bibnamefont {Wang}}, \bibinfo
  {author} {\bibfnamefont {L.}~\bibnamefont {Li}}, \ and\ \bibinfo {author}
  {\bibfnamefont {J.}~\bibnamefont {Xiao}},\ }\bibfield  {title} {\enquote
  {\bibinfo {title} {Anti-phase synchronization of two coupled mechanical
  metronomes},}\ }\href@noop {} {\bibfield  {journal} {\bibinfo  {journal}
  {Chaos}\ }\textbf {\bibinfo {volume} {22}},\ \bibinfo {pages} {023146}
  (\bibinfo {year} {2012})}\BibitemShut {NoStop}%
\bibitem [{\citenamefont
  {Bahraminasab}(2007)}]{bahraminasab2007synchronisation}%
  \BibitemOpen
  \bibfield  {author} {\bibinfo {author} {\bibfnamefont {A.}~\bibnamefont
  {Bahraminasab}},\ }\href@noop {} {\emph {\bibinfo {title} {Synchronisation}}}
  (\bibinfo {year} {2007}),\ \bibinfo {note}
  {\url{https://www.youtube.com/watch?v=W1TMZASCR-I}}\BibitemShut {NoStop}%
\bibitem [{\citenamefont {Ikeguchi-Laboratory}(2012)}]{32metronomes}%
  \BibitemOpen
  \bibfield  {author} {\bibinfo {author} {\bibnamefont {Ikeguchi-Laboratory}},\
  }\href@noop {} {\emph {\bibinfo {title} {Synchronization of thirty two
  metronomes}}} (\bibinfo {year} {2012}),\ \bibinfo {note}
  {\url{https://www.youtube.com/watch?v=JWToUATLGzs}}\BibitemShut {NoStop}%
\bibitem [{\citenamefont {MythBusters}(2014)}]{mythbusters2014nsync}%
  \BibitemOpen
  \bibfield  {author} {\bibinfo {author} {\bibnamefont {MythBusters}},\
  }\href@noop {} {\emph {\bibinfo {title} {N-Sync}}} (\bibinfo {year} {2014}),\
  \bibinfo {note}
  {\url{https://www.youtube.com/watch?v=e-c6S6SdkPo}}\BibitemShut {NoStop}%
\bibitem [{\citenamefont {Lepschy}, \citenamefont {Mian},\ and\ \citenamefont
  {Viaro}(1992)}]{lepschy1992feedback}%
  \BibitemOpen
  \bibfield  {author} {\bibinfo {author} {\bibfnamefont {A.~M.}\ \bibnamefont
  {Lepschy}}, \bibinfo {author} {\bibfnamefont {G.}~\bibnamefont {Mian}}, \
  and\ \bibinfo {author} {\bibfnamefont {U.}~\bibnamefont {Viaro}},\ }\bibfield
   {title} {\enquote {\bibinfo {title} {Feedback control in ancient water and
  mechanical clocks},}\ }\href@noop {} {\bibfield  {journal} {\bibinfo
  {journal} {IEEE Transactions on Education}\ }\textbf {\bibinfo {volume}
  {35}},\ \bibinfo {pages} {3--10} (\bibinfo {year} {1992})}\BibitemShut
  {NoStop}%
\bibitem [{\citenamefont {Moon}\ and\ \citenamefont
  {Stiefel}(2006)}]{moon2006coexisting}%
  \BibitemOpen
  \bibfield  {author} {\bibinfo {author} {\bibfnamefont {F.~C.}\ \bibnamefont
  {Moon}}\ and\ \bibinfo {author} {\bibfnamefont {P.~D.}\ \bibnamefont
  {Stiefel}},\ }\bibfield  {title} {\enquote {\bibinfo {title} {Coexisting
  chaotic and periodic dynamics in clock escapements},}\ }\href@noop {}
  {\bibfield  {journal} {\bibinfo  {journal} {Phil Trans R Soc A}\ }\textbf
  {\bibinfo {volume} {364}},\ \bibinfo {pages} {2539--2564} (\bibinfo {year}
  {2006})}\BibitemShut {NoStop}%
\bibitem [{\citenamefont {Roup}\ \emph {et~al.}(2003)\citenamefont {Roup},
  \citenamefont {Bernstein}, \citenamefont {Nersesov}, \citenamefont {Haddad},\
  and\ \citenamefont {Chellaboina}}]{roup2003limit}%
  \BibitemOpen
  \bibfield  {author} {\bibinfo {author} {\bibfnamefont {A.~V.}\ \bibnamefont
  {Roup}}, \bibinfo {author} {\bibfnamefont {D.~S.}\ \bibnamefont {Bernstein}},
  \bibinfo {author} {\bibfnamefont {S.~G.}\ \bibnamefont {Nersesov}}, \bibinfo
  {author} {\bibfnamefont {W.~M.}\ \bibnamefont {Haddad}}, \ and\ \bibinfo
  {author} {\bibfnamefont {V.}~\bibnamefont {Chellaboina}},\ }\bibfield
  {title} {\enquote {\bibinfo {title} {{Limit cycle analysis of the verge and
  foliot clock escapement using impulsive differential equations and
  Poincar{\'e} maps}},}\ }\href@noop {} {\bibfield  {journal} {\bibinfo
  {journal} {International Journal of Control}\ }\textbf {\bibinfo {volume}
  {76}},\ \bibinfo {pages} {1685--1698} (\bibinfo {year} {2003})}\BibitemShut
  {NoStop}%
\bibitem [{\citenamefont {Rowlings}(1944)}]{rowlings1944science}%
  \BibitemOpen
  \bibfield  {author} {\bibinfo {author} {\bibfnamefont {A.}~\bibnamefont
  {Rowlings}},\ }\href@noop {} {\emph {\bibinfo {title} {The Science of Clocks
  and Watches}}}\ (\bibinfo  {publisher} {Caldwell Industries, Luling, TX
  USA},\ \bibinfo {year} {1944})\BibitemShut {NoStop}%
\bibitem [{\citenamefont {Guckenheimer}\ and\ \citenamefont
  {Holmes}(1983)}]{guckenheimer1983nonlinear}%
  \BibitemOpen
  \bibfield  {author} {\bibinfo {author} {\bibfnamefont {J.}~\bibnamefont
  {Guckenheimer}}\ and\ \bibinfo {author} {\bibfnamefont {P.}~\bibnamefont
  {Holmes}},\ }\href@noop {} {\emph {\bibinfo {title} {Nonlinear Oscillations,
  Dynamical Systems, and Bifurcations of Vector Fields}}}\ (\bibinfo
  {publisher} {Springer},\ \bibinfo {year} {1983})\BibitemShut {NoStop}%
\bibitem [{\citenamefont {Strogatz}(1994)}]{strogatz1994nonlinear}%
  \BibitemOpen
  \bibfield  {author} {\bibinfo {author} {\bibfnamefont {S.~H.}\ \bibnamefont
  {Strogatz}},\ }\href@noop {} {\emph {\bibinfo {title} {Nonlinear Dynamics and
  Chaos}}}\ (\bibinfo  {publisher} {Addison-Wesley},\ \bibinfo {year}
  {1994})\BibitemShut {NoStop}%
\bibitem [{\citenamefont {Bender}\ and\ \citenamefont
  {Orszag}(1999)}]{bender1999advanced}%
  \BibitemOpen
  \bibfield  {author} {\bibinfo {author} {\bibfnamefont {C.~M.}\ \bibnamefont
  {Bender}}\ and\ \bibinfo {author} {\bibfnamefont {S.~A.}\ \bibnamefont
  {Orszag}},\ }\href@noop {} {\emph {\bibinfo {title} {Advanced Mathematical
  Methods for Scientists and Engineers}}}\ (\bibinfo  {publisher} {Springer},\
  \bibinfo {year} {1999})\BibitemShut {NoStop}%
\bibitem [{\citenamefont {Holmes}(1995)}]{holmes1995introduction}%
  \BibitemOpen
  \bibfield  {author} {\bibinfo {author} {\bibfnamefont {M.~H.}\ \bibnamefont
  {Holmes}},\ }\href@noop {} {\emph {\bibinfo {title} {Introduction to
  Perturbation Methods}}}\ (\bibinfo  {publisher} {Springer},\ \bibinfo {year}
  {1995})\BibitemShut {NoStop}%
\bibitem [{\citenamefont {Dhooge}\ \emph {et~al.}(2008)\citenamefont {Dhooge},
  \citenamefont {Govaerts}, \citenamefont {Kuznetsov}, \citenamefont {Meijer},\
  and\ \citenamefont {Sautois}}]{dhooge2008new}%
  \BibitemOpen
  \bibfield  {author} {\bibinfo {author} {\bibfnamefont {A.}~\bibnamefont
  {Dhooge}}, \bibinfo {author} {\bibfnamefont {W.}~\bibnamefont {Govaerts}},
  \bibinfo {author} {\bibfnamefont {Y.~A.}\ \bibnamefont {Kuznetsov}}, \bibinfo
  {author} {\bibfnamefont {H.~G.~E.}\ \bibnamefont {Meijer}}, \ and\ \bibinfo
  {author} {\bibfnamefont {B.}~\bibnamefont {Sautois}},\ }\bibfield  {title}
  {\enquote {\bibinfo {title} {New features of the software matcont for
  bifurcation analysis of dynamical systems},}\ }\href@noop {} {\bibfield
  {journal} {\bibinfo  {journal} {Mathematical and Computer Modelling of
  Dynamical Systems}\ }\textbf {\bibinfo {volume} {14}},\ \bibinfo {pages}
  {147--175} (\bibinfo {year} {2008})}\BibitemShut {NoStop}%
\bibitem [{\citenamefont {Fradkov}\ and\ \citenamefont
  {Andrievsky}(2007)}]{fradkov2007synchronization}%
  \BibitemOpen
  \bibfield  {author} {\bibinfo {author} {\bibfnamefont {A.~L.}\ \bibnamefont
  {Fradkov}}\ and\ \bibinfo {author} {\bibfnamefont {B.}~\bibnamefont
  {Andrievsky}},\ }\bibfield  {title} {\enquote {\bibinfo {title}
  {Synchronization and phase relations in the motion of two-pendulum system},}\
  }\href@noop {} {\bibfield  {journal} {\bibinfo  {journal} {International
  Journal of Non-Linear Mechanics}\ }\textbf {\bibinfo {volume} {42}},\
  \bibinfo {pages} {895--901} (\bibinfo {year} {2007})}\BibitemShut {NoStop}%
\bibitem [{\citenamefont {Kumon}\ \emph {et~al.}(2002)\citenamefont {Kumon},
  \citenamefont {Washizaki}, \citenamefont {Sato}, \citenamefont {Mizumoto},\
  and\ \citenamefont {Iwai}}]{kumon2002controlled}%
  \BibitemOpen
  \bibfield  {author} {\bibinfo {author} {\bibfnamefont {M.}~\bibnamefont
  {Kumon}}, \bibinfo {author} {\bibfnamefont {R.}~\bibnamefont {Washizaki}},
  \bibinfo {author} {\bibfnamefont {J.}~\bibnamefont {Sato}}, \bibinfo {author}
  {\bibfnamefont {R.}~\bibnamefont {Mizumoto}}, \ and\ \bibinfo {author}
  {\bibfnamefont {Z.}~\bibnamefont {Iwai}},\ }\bibfield  {title} {\enquote
  {\bibinfo {title} {{Controlled synchronization of two 1-DOF coupled
  oscillators}},}\ }in\ \href@noop {} {\emph {\bibinfo {booktitle} {Proceedings
  of the 15th IFAC World Congress, Barcelona}}}\ (\bibinfo {year} {2002})\ pp.\
  \bibinfo {pages} {3--10}\BibitemShut {NoStop}%
\end{thebibliography}%

\end{document}